%% file: main.tex
\documentclass[sigconf, nonacm]{acmart}
\usepackage{latexsym}
\usepackage{soul}
\usepackage{graphicx} % DO NOT CHANGE THIS
\usepackage{multirow}
\usepackage{xcolor}
\usepackage{tabularx}
\usepackage{booktabs}
\usepackage{balance}
\usepackage[rightcaption]{sidecap}
\usepackage{microtype}
\usepackage{url}
\usepackage{diagbox}
\usepackage{subcaption}
\usepackage{dsfont}
\usepackage{etoolbox}
\usepackage[linesnumbered,ruled]{algorithm2e}

\newcommand{\up}[1]{\tiny ($\textcolor{green}{\blacktriangle}#1\%)$}
\newcommand{\down}[1]{\tiny ($\textcolor{red}{\blacktriangledown}#1\%)$}

\newcommand{\robust}{\textsc{Robust04}}
\newcommand{\trecdl}{\textsc{TREC-DL}}
\newcommand{\trecdldf}{\textsc{Doc'19}}
\newcommand{\trecdlds}{\textsc{Doc'20}}
\newcommand{\scifact}{\textsc{SciFact}}
\newcommand{\fiqa}{\textsc{FiQA}}
\newcommand{\ms}{\textsc{MS MARCO}}

\newcommand{\bert}{\textsc{BERT}}
\newcommand{\roberta}{\textsc{RoBERTa}}
\newcommand{\distilbert}{\textsc{DistilBERT}}

\newcommand{\bm}{\textsc{BM25}}
\newcommand{\glove}{\textsc{GloVe}}
\newcommand{\random}{\textsc{Sampling}}
\newcommand{\baseline}{\texttt{Baseline}}
\newcommand{\base}{\texttt{Base}}
\newcommand{\ce}{\texttt{Pointwise}}
\newcommand{\pairwise}{\texttt{Pairwise}}
\newcommand{\scl}{\texttt{RankingSCL}}

\begin{document}
\title{Supervised Contrastive Learning Approach for Contextual Ranking}

\author{Abhijit Anand}
\affiliation{
  \institution{L3S Research Center}
  \city{Hannover}
  \country{Germany}
}
\email{aanand@L3S.de}

\author{Jurek Leonhardt}
\affiliation{
  \institution{L3S Research Center}
  \city{Hannover}
  \country{Germany}
}
\email{leonhardt@L3S.de}

\author{Koustav Rudra}
\affiliation{
  \institution{Indian Institute of Technology\\(Indian School of Mines)}
  \city{Dhanbad}
  \country{India}
}
\email{koustav@iitism.ac.in}
\authornote{Research was primarily conducted while affiliated to L3S Research Center.}

\author{Avishek Anand}
\affiliation{
  \institution{Delft University of Technology}
  \city{Delft}
  \country{Netherlands}
}
\email{avishek.anand@tudelft.nl}
\authornotemark[1]
\input{abstract}




\maketitle

\input{intro}

\input{related}

\input{method}
\input{evaluation}

\input{conclusion}

\begin{acks}
This work is supported by the European Union – Horizon 2020 Program under the scheme “INFRAIA-01-2018-2019 – Integrating Activities for Advanced Communities”, Grant Agreement n.871042, “SoBigData++: European Integrated Infrastructure for Social Mining and Big Data Analytics” (http://www.sobigdata.eu).
\end{acks}

\newpage
% \small{
\bibliographystyle{ACM-Reference-Format}
\bibliography{reference}
% }
\appendix
\end{document}

%% file: abstract.tex
\begin{abstract}
Contextual ranking models have delivered impressive performance improvements over classical models in the document ranking task.
However, these highly over-parameterized models tend to be data-hungry and require large amounts of data even for fine tuning.

This paper proposes a simple yet effective method to improve ranking performance on \textit{smaller datasets} using \textit{supervised contrastive learning} for the document ranking problem.
We perform data augmentation by creating training data using parts of the relevant documents in the query-document pairs. 
We then use a supervised contrastive learning objective to learn an effective ranking model from the augmented dataset.
Our experiments on subsets of the \trecdl{} dataset show that, although data augmentation leads to an increasing the training data sizes, it does not necessarily improve the performance using existing pointwise or pairwise training objectives. However, our proposed supervised contrastive loss objective leads to performance improvements over the standard non-augmented setting showcasing the utility of data augmentation using contrastive losses. 
Finally, we show the real benefit of using supervised contrastive learning objectives by showing marked improvements in smaller ranking datasets relating to news (\robust{}), finance (\fiqa{}), and scientific fact checking (\scifact{}). 
\end{abstract}

%% file: intro.tex
\section{Introduction}
\label{sec:intro}

Recent approaches for ranking documents have focused heavily on contextual transformer-based models for both retrieval~\cite{lin2019neural,khattab2020colbert} and re-ranking~\cite{dai_sigir_2019,macavaney2019contextualized:bertir:mpi,yilmaz2019cross,hofstatter2020interpretable,li2020parade}.
To further improve the effectiveness of contextual ranking models, earlier works have explored negative sampling techniques~\cite{xiong2020approximate}, pre-training approaches~\cite{chang2020pre}, and different architectural variants \cite{khattab2020colbert, hofstatter2020interpretable}.
In this paper we investigate the use of simple yet effective data augmentation techniques for ad-hoc document retrieval.

\emph{Data augmentation} (DA) encompasses methods of increasing training data without directly collecting more data but by either adding slightly modified copies of existing data or creating synthetic data.
Data augmentation has been successfully used to help train more robust models, particularly when using smaller datasets in computer vision~\cite{shorten_2019_image_augmentation_survey}, speech recognition~\cite{nguyen_2020_speech_aug}, spoken language understanding~\cite{peng_2020_spoken_lang}, and dialog system~\cite{zhu_2020_dialog}. 
Most of the augmentation approaches are based on a heuristic set of rules based on well-understood domain-specific phenomena.  
However, the use of data augmentation for document ranking has not been investigated in detail to the best of our knowledge.

Both in NLP and IR tasks, the use of large amounts of language data to \textit{pre-train} an initial version of an contextual model, followed by refinement or \textit{fine-tuning} using a small amount of domain-specific data this has delivered impressive gains both in sample efficiency and better generalization.
However, popular contextualized models are over-parameterized with more than \textit{100 million} parameters and might overfit the training data when the task-specific fine-tuning data is small.
Training data for rankings can be either small due to smaller query workloads~\cite{robust04} or incomplete labels as in~\cite{nguyen2016ms_marco} and this is where data augmentation techniques can be valuable.
However, simply augmenting training data with existing point- or pairwise ranking losses does not lead to performance improvements.
In fact, we show that our data augmentation techniques using existing pointwise ranking losses, i.e. cross-entropy losses, results in \textit{degradation of performance} in many cases.
This can be attributed to known lack of robustness to noisy labels~\cite{zhang2018generalized::augprob} and the possibility of poor margins~\cite{liu2016large:augprob}, leading to reduced generalization performance.

Towards improving training in the \textit{limited data setting} using data augmentation, we propose \textit{supervised contrastive learning objectives} (\scl{}) for document ranking.
A key idea in contrastive learning is to learn input representation of an instance or \textit{anchor} such that its positive instances are embedded closer to each other and the negative samples are farther apart.
In this work we construct \textbf{augmented query-document} from existing positive instances by multiple augmentation strategies.
We extend the idea of supervised contrastive learning (SCL) to the document ranking task by considering query-document pairs belonging to the same query as positive instances, unlike in vision and NLP tasks, where all instances with the same class label can potentially become positive pairs.
An important technical challenge while extending SCL loss to ranking data is that of \textit{data sparsity of positive pairs}. 
One could, in principle, decrease data sparsity by including multiple positive instances of the query in the same batch. 
However, decreasing sparsity also results in decreased randomness, which is crucial in training generalizable ranking models.
Towards this, we propose a \textit{practical batching strategy} that maximises randomness while allowing for augmented query-document pairs.

We conduct extensive experiments on multiple contextual models -- \bert{}, \roberta{}, \distilbert{} -- and multiple low-data ranking settings to establish the effectiveness of our approaches.
Note that our primary aim in this paper is to improve ranking performance for smaller ranking datasets by using simple data-augmentation techniques. 
We do not intend to engineer a state-of-art ranking model for document ranking, but instead focus on optimization strategies that work well for simple data augmentation techniques in low data settings. 

\subsection{Research Questions}
\label{sec:intro-rqs}
\begin{itemize}
    \item[\textbf{RQ I.}] 
     Does data augmentation or Supervised Contrastive Learning help to improve document re-ranking performance for smaller datasets?
    
    \item[\textbf{RQ II.}]
    Does the augmentation style impact the performance?
    
    \item[\textbf{RQ III.}]
    How does training data size impact performance?
    
\end{itemize}

Towards answering these research questions we conduct extensive experiments on the following ranking datasets -- \trecdl{}, \robust{}, \fiqa{}, and \scifact{}. 
While \trecdl{} and \robust{} contain longer documents and under-specified queries, \fiqa{} is a question-answering dataset over financial text, and \scifact{} deals with fact checking queries.
Note that \fiqa{} and \scifact{} are much smaller datasets compared to \trecdl{}.  

\subsection{Contributions}
%\label{sec:intro-rqs}

In sum, we make the following contributions in this work:

\begin{itemize}
    \item 
    We propose and study data augmentation approaches for document ranking.
    
    \item
    We propose a ranking-based supervised contrastive loss for exploiting positive augmented pairs for improving ranking performance.
    
    \item
    We show that our ranking-based SCL delivers substantial performance improvements for a wide variety of ranking models under both low and high data settings.
\end{itemize}

%% file: related.tex
\section{Related Work}
\label{sec:related}
The related work can be broadly divided into three areas. We start by outlining prominent strategies for document ranking using contextual models.
Next, we review various data augmentation strategies and their application in text tasks.  
Finally we reflect upon various loss functions used in text ranking and their relationship with supervised contrastive loss.

\subsection{Contextual Models for Ad-hoc Document Retrieval}
\label{sec:rel-work:contextual}

A standard strategy for the text ranking task involves a fast retrieval step followed by a more involved \textit{re-ranking} step.
In this paper, we are concerned with  improving the performance of the  re-ranking stage that typically involves the use of contextual models.
Contextual models like~\cite{devlin_bert_2018,liu2019roberta} have shown promising improvements in document ranking task~\cite{dai_sigir_2019,rudra2020distant,li2020parade}. 

There are two major paradigms to encode the input, i.e., query document pairs, for training a contextual re-ranker -- (a) joint encoding, and (b) independent encoding.
The most common way for applying contextual models for the problem of document re-ranking is to \textbf{jointly encode} the query and document using a over-parameterized language model~\cite{nogueira_prr_2019,macavaney2019contextualized:bertir:mpi}.
Independent encoding, on the other hand, encoding the document and the query independently of each other. 
Such models that implement independent query and document encoding are called \textbf{dual encoders}, \textbf{bi-encoders}, or \textbf{two-tower models}.
Typically, dual encoders are used in the retrieval phase~\cite{lee2019latent,ahmad2019reqa,chang2020pre,karpukhin2020dense,khattab2020colbert,hofstatter2020interpretable} however there have been recent proposals that use dual encoders in the re-ranking phase~\cite{leonhardt2021fast}. 
Note that a common problem in both approaches is due to an upper bound on the acceptable input length of contextual models that restricts its applicability to shorter documents.
When documents do not fit into the model the documents are chunked into passages/sentences to fit within token limit either by using transformer-kernels~\cite{hofstatter2020local,hofstatter2020interpretable}, truncation~\cite{dai_sigir_2019}, or careful pre-selection of relevant text~\cite{rudra2020distant,leonhardt2021learnt}. 

In this work, we focus on the joint encoding models for document ranking and employ simple document truncation whenever longer documents overflow the overall input upper bound.

\subsection{Data Augmentation}
Data augmentation has a significant impact in different segments such as text, speech, image, vision, etc. Researchers proposed new data augmentation strategies~\cite{chen_2020_gridmask,cubuk_2019_autoaugment,zhong_2020_erasing} and their influence on deep learning models~\cite{gao_2020_fuzz,raileanu_2020_generalize_reinforce,zeng_2020_adversarial,longpre_2020_effective}. Data augmentation helps in speech recognition~\cite{nguyen_2020_speech_aug}, spoken language understanding~\cite{peng_2020_spoken_lang}, and dialog system~\cite{zhu_2020_dialog}. Data augmentation~\cite{kumar_2020_augment_transformer,sun_2020_mixup,shorten2021text} using pre-trained transformer models show significant boost in the performance of several downstream natural language processing (nlp) and text related tasks. Morris~et~al~\cite{morris_2020_textattack} proposed a framework \textit{Text Attack} for data augmentation, adversarial attacks, and training in nlp. 
Different natural language tasks such as named entity recognition~\cite{liu_mulda_low_resource_aug}, language inference~\cite{dong_2021_cross_nli}, text categorization, classification~\cite{yu_2019_hierarchical,lu_2006_enhancing}, query based multi-document summarization~\cite{pasunuru2021data}. 
Image and vision related tasks also significantly benefit through data augmentation. Shorten~et~al~\cite{shorten_2019_image_augmentation_survey} provides a survey on the role of image data augmentation strategy on deep learning. 
Data augmentation helps to boost performance in multiple image/vision related tasks such as user identification~\cite{mekruksavanich_2021_user_identify}, image retrieval~\cite{yanagi_2020_image}, image segmentation~\cite{xu_2020_image_segment}, text recognition and object detection~\cite{li_2020_recognition,luo_2020_learn_augment,yanagi_2020_image,zoph_2020_object_detection}.

Recently, data augmentation strategies have been deployed for retrieval tasks. It shows promising results for question retrieval~\cite{nugraha_2019_typographic}, query translation~\cite{yao_2020_domain}, question-answering~\cite{yang_2019_data,yang_2020_cross_attention}, cross-language sentence selection~\cite{chen_2021_cross}, machine reading~\cite{van_2021_machine_reading}, query expansion~\cite{lian_2020_retrieve_keyword}. Yang~et~al~\cite{yang2021xmoco} proposed cross-momentum contrastive learning~\cite{he2020momentum} based open-domain question answering scheme. Recent dense retriever models~\cite{xiong2020approximate,karpukhin2020dense} sample negative documents to train dense retrievers in contrastive way. However, such methods do not take care of uniformity nature of contrastive learning~\cite{wang2020understanding}.  Li~et~al~\cite{li_2021_dual_learning} proposed a contrastive dual learning based method for dense retrieval that takes care of uniformity. Most of these strategies focus on negative samples and try to train an efficient dense retriever framework. Data augmentation strategy with contrastive loss setup is also not yet explored for document ranking task. In this paper, we take a step towards that and explore the effect of different \textit{data augmentation strategies} with \textit{supervised contrastive learning} setup on the re-ranking performance.

\subsection{Supervised Contrastive Learning (SCL)}
\label{sec:rel-work:scl}

Contrastive losses using data augmentation have been popularized in the machine learning literature in the unsupervised learning setting. Specifically, augmentation of an instance are treated as positive samples and other random instances from the batch are treated as negative samples. We are inspired from the recent idea of contrastive loss that also exploit the label information for more fine-grained supervision signal from data augmentation~\cite{khosla2020supervised}. 
Recent methods utilize this approach to learn representations from unsupervised data~\cite{oord2018representation,hjelm2018learning,wu2018unsupervised,sohn2016improved} and they outperform other approaches~\cite{donahue2019large,gidaris2018unsupervised}. 
Training instances are generated from original ones using different data augmentation strategies and contrastive loss helps to bring the representation of similar/related entities close to each other in the embedding space. 
For a more detailed overview we point the interested reader to a recent survey on supervised and self-supervised  contrastive learning~\cite{jaiswal2021survey}. 
Recently, SCL has been applied to  fine-tuning regimes using pre-trained language models but with limited success~\cite{gunel2020supervised}, and also for the retrieval stage (not re-ranking)~\cite{li2021more}. 
To the best of our knowledge, SCL has not been used in document ranking using joint encoder models.

The learning objective of neural ranking models is broadly studied under three types -- pointwise, pairwise, and list-wise losses. 
A pointwise learning objective tries to optimize a ranking model by trying to directly predict the relevance class using for example the widely popular cross-entropy loss.
Pairwise ranking objectives, on other hand, focus on optimizing the  preference-pairs induced by the document labels in the training dataset.
Note that when using pairwise losses, the aim is to always distinguish between different labels, i.e., relevant vs irrelevant or highly relevant vs relevant. 
Finally list-wise losses directly try to optimize the ranking as a whole.
In principle contrastive losses can be used in conjugation with any of the aforementioned losses and in this paper we experiment with pointwise and pairwise losses.

The idea of supervised contrastive loss has its roots in self-supervised contrastive learning.

%% file: method.tex
\section{Method}
\label{sec:method}
In this section we begin by defining the document re-ranking problem (cf. Section \ref{sec:problem}). We then describe our contributions, which comprise multiple training data augmentation techniques for re-ranking data (Section \ref{sec:batch_creation}) and a supervised contrastive learning objective which is used to train our ranking models (Section \ref{sec:scl}).

\subsection{Contextual Document Rankers}
\label{sec:problem}
In this paper we aim to learn a \textit{document re-ranking} model. Given a query-document pair $(q, d)$ as input, the model outputs a \textit{relevance score}. This relevance score may then be used to rank a number of documents with respect to their relevance to a given query.

Formally, we have a training set of pairs $\{q_i, d_i \}_{i=1}^{N}$, where $q_i$ is a query and $d_i$ is a document that is either relevant or irrelevant to it, depending on its label $y_i$. Our goal is to train a ranker $R$ that predicts a relevance score $\hat{y} \in [0; 1]$ given a query $q$ and a document $d$:
\begin{equation}
    R: (q, d) \mapsto \hat{y}
\end{equation}
Finally, the trained ranking model $R$ can be used to re-rank a set of documents obtained in the first-stage retrieval process by a lightweight, typically term-frequency-based, retriever w.r.t.\ a query. This is the usual practice for ranking tasks, where the documents are retrieved first and then re-ranked by a more involved and computationally expensive model. In recent times, pre-trained contextual language models have shown promising performance for document ranking task~\cite{nogueira_prr_2019,dai_sigir_2019,rudra2020distant,yilmaz2019cross}. Such cross-attention models jointly model queries and documents. In this paper, we consider three different joint modeling approaches based on \bert{}~\cite{devlin_bert_2018}, \roberta{}~\cite{liu2019roberta} and \distilbert{}~\cite{sanh2019distilbert} and check their performance under supervised contrastive learning setup with different amount of data augmentation. All three models share the same input format: a pair of query $q$ and document $d$ is fed into the model as
\begin{equation}
    \texttt{[CLS]}\ q\ \texttt{[SEP]}\ d\ \texttt{[SEP]}.
\end{equation}
Due to the input length limitation of the models, long documents may be truncated to fit.

\subsection{Supervised Contrastive Learning for Rankings}
\label{sec:scl}
For training, we operate in the mini-batch training setup with a batch of training examples $\left\{x_i, y_i\right\}_{i=1, ..., N}$. Traditionally, ranking models are often trained in one of the following ways:

In \textbf{pointwise} training, the document ranking task is considered as a binary classification problem with a relevant and a non-relevant class. Each training instance $x_i = (q_i, d_i)$ is a query-document pair and $y_i \in \{0, 1\}$ is a relevance label. Let $\hat{y}_i$ be the predicted score of $x_i$. The cross-entropy loss function can be written as follows:
\begin{equation}
    \mathcal{L}_{\mathtt{Point}} = -\frac{1}{N} \sum_{i=1}^{N} \left( y_{i} \cdot \log \hat{y}_{i} + (1 - y_{i}) \cdot \log (1-\hat{y}_{i}) \right)
\end{equation}

In \textbf{pairwise} training, each training example consists of a query and two documents, i.e.\ $x_i = (q_i, d^+_i, d^-_i)$, where the former is more relevant to the query than the latter. The pairwise loss function is
\begin{equation}
    \mathcal{L}_{\mathtt{Pair}} = \frac{1}{N} \sum_{i=1}^{N} \max \left\{0, m - \hat{y}^+_i + \hat{y}^-_i \right\},
\end{equation}
where $\hat{y}^+_i$ and $\hat{y}^-_i$ are the predicted scores of $d^+_i$ and $d^-_i$, respectively, and $m$ is the \emph{loss margin}.

We propose a novel ranking objective that includes a \textbf{supervised contrastive learning} (SCL) term for fine-tuning contextual ranking models in addition to the standard ranking loss. The SCL loss is meant to capture the similarities between relevant parts of documents for the same query and contrast them with the examples from non-relevant queries. Let $\Phi(\cdot) \in \mathbb{R}^{t}$ denote the query-document representation that is output by the ranking model (for example, this corresponds to the \texttt{[CLS]} output for \bert{}-based models).
Let $N_{+}$ be the total number of positive examples in the batch (relevant query-document pairs). $\tau>0$ is an adjustable scalar temperature parameter that controls the separation between the relevant and non-relevant examples and $\lambda$ is a scalar weighting hyper-parameter that we tune for each downstream task and setting. The SCL loss can be written as
\begin{multline}
    \mathcal{L}_{\mathtt{SCL}} = \\
        \sum_{i=1}^{N}-\frac{1}{N_{+}}
        \sum_{j=1}^{N_{+}}
        \mathbf{1}_{\substack{q_i = q_j, \\
                    i \neq j, \\
                    y_i = y_j = 1}}
        \log \frac  {\exp \left(\Phi \left(x_{i} \right) \cdot \Phi \left(x_{j} \right) / \tau \right)}
                    {\sum_{k=1}^{N}
                        \mathbf{1}_{i \neq k} \exp \left(\Phi \left(x_{i} \right) \cdot \Phi \left(x_{k} \right) / \tau \right)}
\end{multline}
%where $q(x_i)$ is the query identifier corresponding to $x_i$, i.e.\ $q(x_i) = q_i$.
Note that $\mathcal{L}_{\mathtt{SCL}}$ constrains the positive pair that has the same query to be embedded close to each other instead of a pair of documents that are relevant for different queries. This is crucial since we want to enforce that the representations for the ``relevant parts'' of the same query be close to each other.

\begin{figure}
    \centering
    \includegraphics[width=\columnwidth]{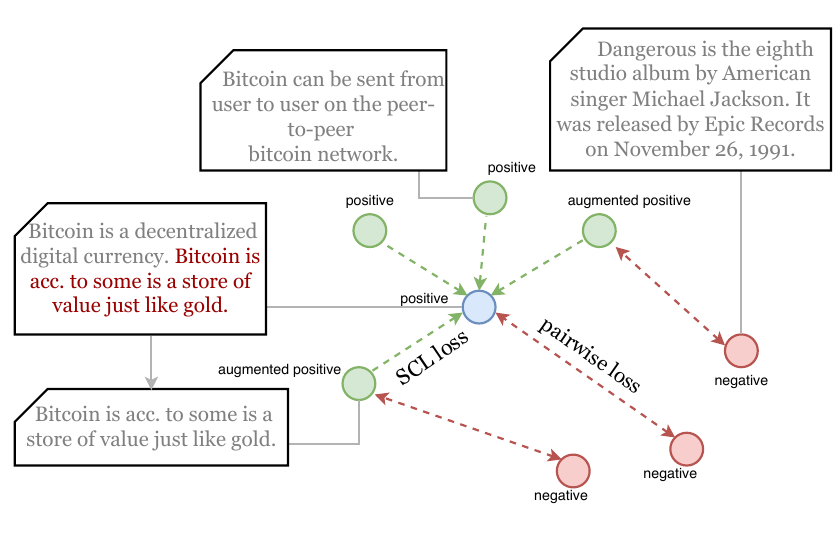}
    \caption{Example batch using data augmentation for rankings for the example query \texttt{bitcoin currency}. Augmented positives are derived from a relevant document, negatives are randomly sampled from the batch. SCL loss tries to bring the representations of positives closer, while pairwise loss repels the representations of positives and negatives apart.}
    \label{fig:bitcoin}
\end{figure}

The overall ranking SCL loss is then given by
\begin{equation}
\label{eqn:scl_equation}
\mathcal{L}_{\mathtt{RankingSCL}} = (1-\lambda) \mathcal{L}_{\mathtt{Ranking}} + \lambda \mathcal{L}_{\mathtt{SCL}},
\end{equation}
where $\mathcal{L}_{\mathtt{Ranking}}$ may be either $\mathcal{L}_{\mathtt{Point}}$ or $\mathcal{L}_{\mathtt{Pair}}$. We illustrate the \scl{} loss in Figure \ref{fig:bitcoin} using a pairwise ranking loss. It shows the two components working together; the ranking loss separates the pairs of positive and negative documents, while the contrastive loss moves all positive documents in the batch closer to each other. We use the following terminology in the paper: linear interpolation of \ce{} and \scl{} is referred to as \textbf{\ce{} \scl{}} and linear interpolation of \pairwise{} and \scl{} is referred to as \textbf{\pairwise{} \scl{}}.

\subsection{Creating and Augmenting Training Batches}
\label{sec:batch_creation}
During the creation of mini-batches, our objective is to preserve the randomness in the data while augmenting the training set. Previous studies showed that the performance of self-supervised contrastive learning depends on the quality of the augmented data~\cite{gunel2020supervised}. 
% The steps involved in the creation of batches are described below:

We start with the top-$k$ documents per query retrieved using a first stage retrieval method. We create the training dataset by collecting all positive query-document training instances from this top-$k$ set. We then randomly sample one irrelevant document for each positive pair. The resulting training set of $(q, d^+, d^-)$ triples is shuffled to ensure randomness. Note that, for pointwise training, we create two query-document pairs from each triple.

\subsubsection{\textbf{Data Augmentation}}
Next, we augment the training instances. For each triple $(q, d^+, d^-)$ in the training set, we create an augmented version $d^+_a$ of $d^+$ by selecting relevant sentences with respect to the corresponding query and randomly sample an irrelevant document $d^-_a$ from the corpus. The augmented training instances are appended to the respective batch. Consequently, after augmentation, each batch contains twice as many training instances. The augmentation approach is illustrated in Algorithm~\ref{alg:augmentation}.

\begin{algorithm}[t]
    \DontPrintSemicolon
    \SetKwFunction{Augment}{augment}
    \SetKw{In}{in}
    \KwIn{training batch $B$}
    \KwOut{augmented training batch $B'$}
    $B' \leftarrow$ empty list\;
    \ForEach{$(q, d^+, d^-)$ \In $B$}{
        \tcp{keep the original example}
        append $(q, d^+, d^-)$ to $B'$\;
        \tcp{create augmented example}
        $d^+_a \leftarrow \Augment(d^+, q)$\;
        $d^-_a \leftarrow$ random irrelevant document\;
        append $(q, d^+_a, d^-_a)$ to $B'$\;
    }
    \Return{$B'$}\;
    \caption{Training data augmentation}
    \label{alg:augmentation}
\end{algorithm}

In order to augment a document, we consider it as a sequence of sentences $s_i$, i.e.\ $d = (s_1, s_2, ..., s_{|d|} )$. A query-specific selector selects a fixed number of sentences from the document. The selector defines a distribution $p(s | q, d)$ over sentences in $d$ given the input query $q$, encoding the relevance of the sentence given the query. This distribution is used to select an extractive, query-dependent summary $d' \subseteq d$. Here, we extract a summary as the 20 highest scoring sentences.

We consider the following three sentence selection strategies:
\begin{itemize}
    \item \textbf{Embedding-based (\glove{})}: We use semantic (cosine) similarity scores between the query $q$ and sentences $s_i$ to determine the best sentences. Both the query and sentence are represented as average over the constituent word embeddings.
    \item \textbf{Term-matching-based (\bm{})}: We use tf-idf scores between the query $q$ and sentences $s_i$ to determine the best sentences. Inverse document frequencies are computed over the complete corpus.
    \item \textbf{Sampling-based (\random{})}: We randomly sample $k$ sentences from the document.
\end{itemize}

%% file: evaluation.tex
\begin{figure*}
    \centering
      
      \includegraphics[width=0.8\linewidth]{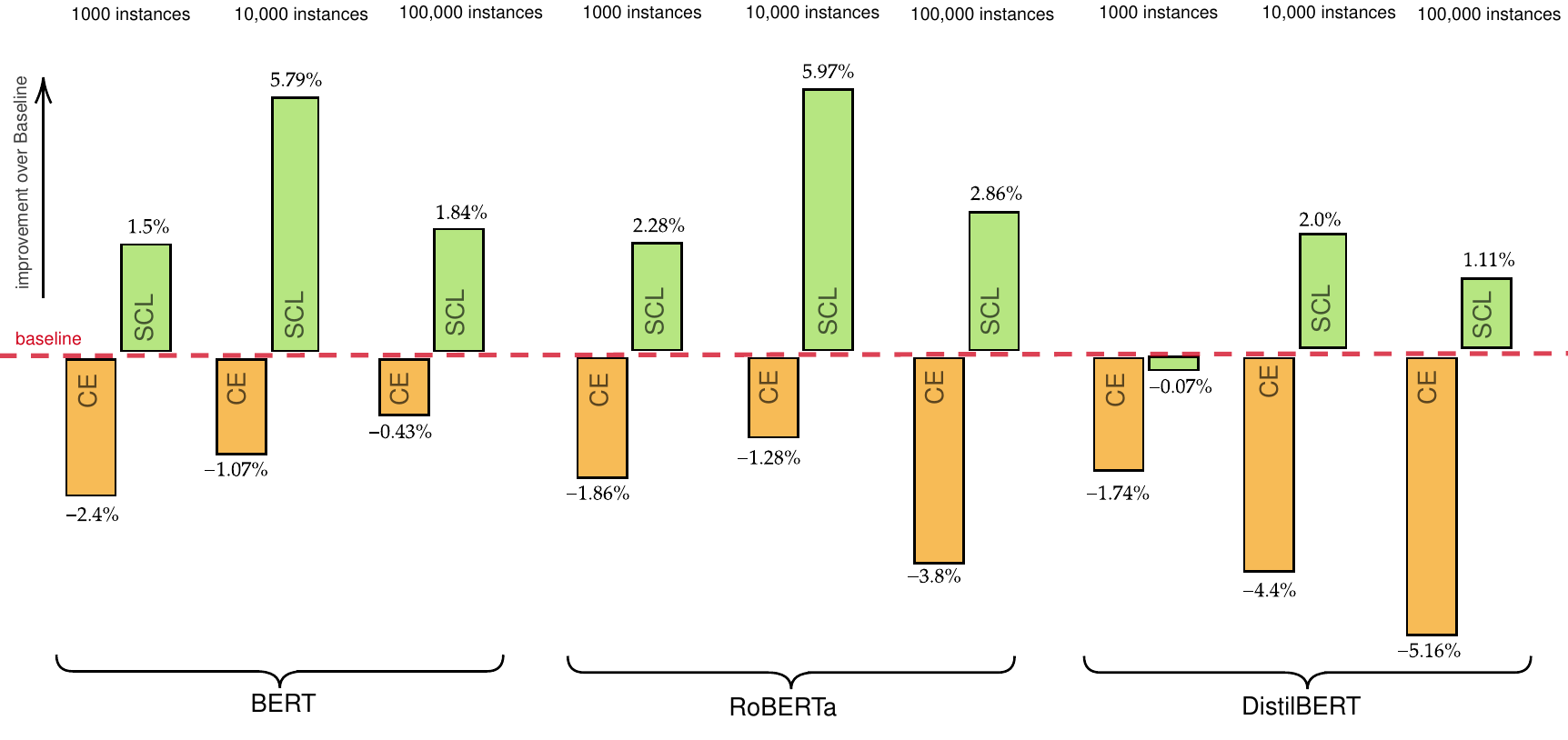}
      \vspace{-5mm}
    \caption{Relative ranking improvement over pointwise baseline ranker trained over non-augmented data. CE in the figure refers to cross entropy (pointwise loss), and SCL refers to pointwise variant of \scl{}.   
    Dataset: \trecdldf{}, augmentation strategy: BM25 selection.}
    \label{fig:ce-scl}
\end{figure*}

\section{Experimental Setup}
In our experiments we answer the following research questions:
\begin{itemize}
    \item[\textbf{RQ I.}] 
     Does data augmentation or Supervised Contrastive Learning help to improve document re-ranking performance for smaller datasets?
    
    \item[\textbf{RQ II.}]
    Does the augmentation style impact the performance?
    
    \item[\textbf{RQ III.}]
    How does training data size impact performance?
    
\end{itemize}

Towards answering these research questions we employ the following datasets, rankers and training settings. Note that we focus on the re-ranking task and not the retrieval task. 

\paragraph{\textbf{Datasets}}
We conduct experiments on the following ranking datasets:
\begin{enumerate}
    \item \textbf{\trecdl{}}: We consider the dataset from the TREC Deep Learning track (2019). We evaluate our model on \trecdldf{} and \trecdlds{} containing 200 queries each. For training and dev set we use \ms{} which contains 367K queries. Top 100 documents are retrieved for each query using Indri~\cite{msmarco_trec_2019}.  
    \item \textbf{\robust{}}: We have $249$ queries with their description and narratives. Along with queries, we also have a $528K$ document collection. Top 100 documents are retrieved for each query using Indri~\cite{strohman_indri_2005} framework. We consider the folds and top documents retrieved directly from~\citet{dai_sigir_2019}.
    \item \textbf{\fiqa{}} was released in 2018 as an open challenge in the Web Conference.\footnote{\url{https://sites.google.com/view/fiqa/home}} It comprises questions and answers from the financial domain, with one of two tasks being \textit{opinion-based QA over financial data}. The QA test collection was crawled from Stackexchange, Reddit and StockTwits. We have in total 6650 queries of which 650 are present in test set and 5500 in training set. The corpus size is around 57K. The top-100 documents are retrieved for each query using \bm{}.
   
    \item \textbf{\scifact{}} \cite{Wadden2020FactOF} is a \emph{scientific fact verification} dataset. We have 1110 queries of which 810 are in the training set and 300 in test set with a document corpus size of 5K. We retrieve the top-100 documents per query using \bm{}. The dataset contains scientific claims written by experts as well as annotated abstracts that may support or refute a given claim. We treat the fact verification task as a ranking problem by retrieving relevant documents for a given query (fact) from the whole corpus.
    
\end{enumerate}

\paragraph{\textbf{Ranking Models}}
We use three different cross-attention models for our experiments:
\begin{enumerate}
    \item \textbf{\bert{}} \cite{devlin_bert_2018} is a large, pre-trained contextual model based on the transformer architecture. We use the \emph{base} version with 12 encoder layers, 12 attention heads and 768-dimensional output representations. The input length is restricted to a maximum of 512 tokens.
    \item \textbf{\roberta{}} \cite{liu2019roberta} is another cross-attention model which is architecturally identical to \bert{}; the only difference of the two models is the pre-training procedure. 
    \item \textbf{\distilbert{}} \cite{sanh2019distilbert} employs \emph{knowledge distillation} techniques to compress the original \bert{} model to roughly 40\% of its original size, while largely maintaining performance. As a result, \distilbert{} is a much smaller parametric ranker with only $6$ encoder layers. We choose \distilbert{} to study the effect of \scl{} and augmentation on models with low parameterization.
\end{enumerate}

\input{tables/table-selectors}

\subsection{Experiments Conducted}
To answer the above RQ's, we experiment with (a) different types of contextual models – BERT, RoBERTa, DistilBERT, (b) varying dataset sizes – {1K, 2K, 10K, 100K} instances for \trecdldf{} and \trecdlds{}, (c) two ranking losses – \ce{} \scl{} and \pairwise{} \scl{} d) different data augmentation strategies \bm{}, \glove{}, \random{} and e) different datasets \trecdldf{}, \trecdlds{}, \robust{}, \scifact{} and \fiqa{}. To give an example, the number of models trained on \trecdldf{} is 1440 (72 best model combinations are chosen for reporting). Given the large number of models it is difficult to report all combination of results and their respective hyperparameters. So we chose to report a selective part of it due to space constraints.

\subsection{Batch Creation and Hyperparameters}
\label{sec:batch-creation}

    As mentioned in Section~\ref{sec:batch_creation}, we consider the positive query document pairs from the top-$k$ retrieved set and sample an equal number of negative pairs for the original dataset. After that, we use the selector to generate augmented versions of documents. For \trecdl, we tried with varying amounts of query-document pairs - 1k, 2k, 10k, and 100k. For example, for 1k, we have 500 positive and 500 negative pairs that constitute our original dataset. Further, we add 1k more through the augmentation process. Hence, 1k contains a total of 2k query-document pairs. The same pattern holds for the other three sizes. In \robust, we consider all the pairs from the training set because it contains fewer queries. \textit{Note that we only augment the training data, the validation and test sets are not augmented.}

\paragraph{\textbf{Hyperparameters}}
We have two hyperparameters in our models: the temperature ($\tau$), and the degree of interpolation ($\lambda$) as in \scl{} [eq. \eqref{eqn:scl_equation}]. 
We use the \ms{} development set to determine the best combination of $\tau$ and $\lambda$. For \robust{}, we use the validation set as shared by \citet{dai_sigir_2019}, i.e.\ a small subset of training queries. These parameters are different for different ranking models and augmentation strategies (\bm{}, \glove{}, \random{}). For example, in \trecdl{}, BERT ranking model using \bm{} data augmentation and \ce{}\scl{} loss objective returns best score on validation set at $\tau=0.4$ and $\lambda=0.8$ values respectively. In all our experiments we use a batch size of 16. Reporting all hyper-parameter values is not possible owing to the large number of experiments.

\section{Experimental Results}
\label{sec:results}

We start by first answering the question if existing loss functions are sufficient in delivering performance improvements when considering augmented datasets.
Next, we take a detailed look into the effect of the training data size, contextual model type and the augmentation strategy on the ranking performance when using \scl{}.
Finally, we look at the impact of \scl{} on training contextual ranking models on smaller datasets.

\subsection{Augmentation with and without SCL}
\label{sec:results-rankingscl}

To answer \textbf{RQ I}, we first compare the performance of rankers trained using the standard loss functions in comparison to Ranking-SCL losses.
We conducted several experiments to check the relative improvement of rankers trained with \ce{}\scl{} and \pairwise{}\scl{} loss on the \textit{same augmented datasets} over different training set sizes and augmentation strategies.
Figure~\ref{fig:ce-scl} plots the relative performance improvement (in terms of MAP) of models trained using data augmentation over a baseline model trained without data augmentation. Note that performance in terms of ranking metrics and augmentation strategies for \ce{}\scl{} and \pairwise{}\scl{} follows a similar pattern and we use Figure~\ref{fig:ce-scl} as a representative result.

\input{tables/table-bm25}
\input{figures}

We observe that the \ce{} loss on the augmented datasets in fact performs worse than the baseline non-augmented variant.
This is not surprising, since it has been shown in the ML literature~\cite{dodge2020fine,zhang2015character} that cross-entropy is sensitive to label and data noise. 

On the other hand, Figure~\ref{fig:ce-scl} shows that \ce{}\scl{} effectively utilizes augmented data to learn better representations.
This is reflected in consistent improvements over the baselines.
By considering increasing amount of training data, i.e.\ 1000 to 100,000 instances, we observe that for \roberta{} and \distilbert{} more data augmentation can negatively impact ranking performance when \scl{} loss is not used.
This establishes that increasing data augmentation with traditional ranking loss functions is detrimental to ranking performance. 

\paragraph{\textbf{Insight 1.}} Our first insight is that data augmentation is useful only when a proper loss function is used in conjugation, i.e.\ \ce\scl{} or \pairwise{}\scl{} loss.

\subsection{The Impact of augmentation type}
\label{sec:results-augmentation-type}

We now answer \textbf{RQ II} by comparing different data augmentation approaches -- \textit{matching-based}, \textit{embeddings-based}, and \textit{sampling-based} augmentation methods.
We show the performance of our data augmentation techniques applied to re-ranking models in Table~\ref{tab:performance_augmentation} on the three datasets that involve longer documents.
This is due to the fact that the likelihood of getting an unrelated piece of text as an augmentation candidate is higher for longer texts.

We see that there are no clear winners. 
Firstly, matching-based augmented datasets result in consistent performance over all datasets and rankers.
Secondly, \random{} augmentation already helps in improving ranking performance with the \ce{}\scl{} loss.
Note that an artifact of the the \trecdldf{} and \trecdlds{} datasets is that most of the queries have exactly one relevant document even if there are multiple relevant documents due to the data collection strategy, i.e.\ both the datasets have incomplete labels. 
Arguably, just having one positive document-per-query results in augmented instances being parts of the original relevant document and even a random selection having a high likelihood of being relevant.

The \robust{} dataset, unlike \trecdldf{} and \trecdlds{}, has multiple documents-per-query with positive relevance labels. Having multiple relevant documents per query results in multiple positive-document pairs without resorting to augmentation.

Interestingly, we see that \random{} is as competitive as \glove{} and \bm{}, even for the case where the labelling is complete, i.e.\ in case of the \robust{} dataset.
The conclusion that we draw from this experiment is that for the common low-data scenarios of smaller instances and incomplete labels simple augmentation approaches like \random{} already provide reasonable improvements when using \ce{}\scl{} as the optimization objective. Experiments with \pairwise{} \scl{} have similar trend to \ce{}\scl{} as described above.

\paragraph{\textbf{Insight 2.}} We find that choice of simple data augmentation strategies do not have a big impact on the ranking performance when using \scl{} (\pairwise{} or \ce{}).

\input{tables/table-beir-robust-results}
\subsection{The Impact of Data Augmentation}
\label{sec:results-augmentation}
%what is our experimental setup to validate our hypothesis ?
To answer \textbf{RQ III}, we experiment with (a) different types of contextual models -- \bert{}, \roberta{}, \distilbert{}, (b) varying dataset sizes -- $\{1000, 2000, 10000, 100000 \}$ instances, and (c) two ranking losses -- \ce{}\scl{} and \pairwise{}\scl{} with (d) different data augmentation strategies \bm{}, \glove{}, \random{} -- on a ranking dataset \trecdldf{}. 
In Table~\ref{tab:pointwise-vs-pairwise} we present the results of this experiment where we choose the BM25 augmentation strategy for our three contextual rankers.
We compare the relative ranking improvements of using data augmentation against a baseline that is trained over a non-augmented dataset.
\textit{Note that we refer to the fine-tuned model on the non-augmented dataset as the baseline model.}
Specifically, for a non-augmented dataset (say $1000$,$10000$, or $100000$ instances) an augmented dataset is constructed as described in the previous section (see Section~\ref{sec:batch-creation}). 

The reported results measure the ranking performance when the contextual models are fine-tuned on the augmented dataset using the \texttt{RankingSCL} loss.
The corresponding values in the parentheses measure the increase or decrease in performance compared to the corresponding baseline model (as described earlier).

%What is the main insight and observation and why do you think thats the case ?
In general, we clearly observe that the ranking performance increases in a majority of cases when using data augmentation using the \texttt{RankingSCL} loss function.
Firstly, data augmentation is particularly useful for smaller instances, i.e.\ dataset of sizes $10,000$ instances or less.
Specifically, we see improvements of up to $12.9\%$ in reciprocal rank when using \bert{} ranker (with augmented data) over the baseline \bert{} ranker (without augmentation) in the \trecdldf{} dataset using \ce \scl{}.
To put the query workload into context, note that the \trecdldf{} dataset contains $300,000$ training instances.
The superior performance using augmentation for smaller datasets can be clearly attributed to the small number of training queries, which is insufficient for training over-parameterized contextual rankers without any augmentation.
However, when the training set increases to $100,000$ instances, i.e.\ closer to the full size of the dataset, we see diminishing marginal utility of using data augmentation.
Similar results have also been reported in other studies in NLP~\cite{gunel2020supervised} while fine tuning contextual models for other language tasks.

%What is the next main insight and observation and why do you think thats the case ?
Secondly, we observe that the improvements are much larger when using the \distilbert{} ranker instead of \bert{} or \roberta{} especially in the \pairwise{}\scl{} setting.
We present a grouped bar plot to clearly show the trends in Figure~\ref{fig:trainingset_size}. 
To start off, the \distilbert{} ranker performs poorly in the low-data regime when using both the baseline non-augmented setting as well as in the case of augmentation.
However, when using data augmentation for slightly larger datasets, the performance improvement over the baseline is considerable.
Especially, for the $10,000$ instance dataset, we see an improvement of around $4\%$ in reciprocal rank and $7.7\%$ in NDCG (also statistically significant).
More striking is the performance improvement in the \pairwise{}\scl{} setting, where the NDCG improvements are about $60\%$ for the $1000$ instance dataset.
This shows that for models that require large amounts of training data to perform well, such as \distilbert{}, our training data augmentation techniques turn out to be particularly useful, achieving large improvements over the baseline models and matching or even improving on the other models, despite the rather poor baseline performance.
A similar trend is also seen in \roberta{} for the smallest dataset with performance improvements are considerably large, e.g.\ more than $25\%$ MAP and more than $50\%$ in NDCG for pairwise learning.

\paragraph{\textbf{Insight 3.}} \scl{} has the highest marginal utility when the dataset sizes are small. The utility diminishes with increasing dataset size.

\subsection{The Effect of SCL on Small Datasets}
\label{sec:expt:small-datasets}

A natural question to ask from the last experiment is whether the performance improvements on smaller subsets of \trecdl{} can be replicated on other diverse ranking datasets that have small training data sizes.
In the next experiment, we considered datasets corresponding to three diverse tasks -- a question answering task, a fact verification task, and a document ranking task -- to finally evaluate our claim about the high utility of \scl{} over smaller text ranking datasets.
Both the question-answering task (\fiqa{}) and the fact-verification task (\scifact{}) rank passages given a query that intends to maximize the likelihood of finding the right evidence document at the top part of the ranking.

Since the datasets used in this experiment are much smaller in comparison to the previously used web datasets, we trained multiple contextual models with different initializations and present the average ranking performance results in Table~\ref{tab:robust_beir}.
Note that, variance of the ranking metrics in fine-tuning over smaller datasets using over-parameterized contextual models is a known phenomena~\cite{mosbach2020stability, rogers2020primer, dodge2020fine, wang2020understanding}.
It is clear that, although there is reasonable variance due to model initialization, \scl{} losses result in improved ranking performance, sometimes by a large margin.
To avoid further variance due to the training process and small test set sizes we report average ranking scores over \textit{five runs} as mentioned in~\cite{mosbach2020stability}. 
We observe that both \ce{}\scl{} and \pairwise{}\scl{} result in consistent performance gains for \scifact{} and \fiqa{}.
Impressive improvements are observed when training \bert{} with \pairwise{} \scl{} loss for the\fiqa{}.
This is primarily because the baseline is ineffective to train a reasonable passage ranking model.
In general, \pairwise{} \scl{} outperform \ce{} variants except in \robust{} datasets. One might argue that this is due to the large variance of the baseline when training on smaller datasets. 

\paragraph{\textbf{Insight 4.}} \scl{} results in large performance gains on a variety of small ranking datasets.

\subsection{Threats to Validity}
There are some threats to validity of our work that we detail in the following. Firstly, we put into perspective the actual gains or improvements from our experiments by analyzing if the improvements are statistically significant. We observe an important pattern that is worth discussing. 
The average improvement in the \fiqa{} dataset using \roberta{} when considering \ce{}\scl{} losses is above $100\%$ but interestingly the improvements do not turn out to be statistically significant. 
On the other hand, even if the average improvements in the \pairwise{} \scl{} are lesser than \ce{} \scl{} the improvements turn out to be statistically significant.
On closer examination, it turns out that there is a large variance in the ranking metrics for the \scl{} model when trained in the pointwise regime, i.e., MAP value of $0.24 \pm 0.11$. 
In contrast, the MAP values (with variance) for the baseline model over the test set queries is $0.11 \pm 0.005$ showing the small variance in scores.
This means a small set of queries starkly outperforming the baseline \ce{}  model while there is little difference between a large fraction of queries.
We see a similar pattern in the \bert{} model trained on pairwise \scl{} loss for the \fiqa{} and \robust{}.

%% file: tables/table-selectors.tex
\begin{table*}
    \centering
    % \resizebox{\columnwidth}{!}{
    \begin{tabular}{lp{1.4cm}p{1.2cm}p{1.2cm}p{0.0cm}p{1.4cm}p{1.4cm}p{1.4cm}p{0.0cm}p{1.4cm}p{1.2cm}p{1.4cm}}
    %\begin{tabular}{lccccccccccc}
        \toprule
            & \multicolumn{3}{c}{\trecdldf{}} 
            && \multicolumn{3}{c}{\trecdlds{}}
            && \multicolumn{3}{c}{\robust{}}\\
            \cmidrule(lr){2-4}
            \cmidrule(lr){6-8}
            \cmidrule(lr){10-12}
            & $\text{AP}$ & $\text{RR}$ & $\text{nDCG}_\text{10}$ &
            & 
            $\text{AP}$ & $\text{RR}$ & $\text{nDCG}_\text{10}$ &
            & 
            $\text{AP}$ & $\text{RR}$ & $\text{nDCG}_\text{10}$ \\
        \midrule
        \multicolumn{6}{l}{\bf \bert{}} \\
        \baseline       &   0.244 &  0.834  & 0.592   && 0.373   & 0.891   & 0.547 && 0.264   &  0.763  & 0.506 \\
        \random       & 0.253\up{3.8}   & 0.886\up{6.3}   & 0.617\up{4.3}   &  & 0.391\up{4.8}   & 0.941\up{5.6}   & 0.594\up{8.6}$^{*}$ & & 0.276\up{4.7}   & 0.797\up{4.5}   & 0.537\up{6} \\
\bm           & 0.258\up{5.8}   & 0.924\up{10.8}   & 0.617\up{4.3}   &    & 0.378\up{1.3}   & 0.944\up{6.0}   & 0.562\up{2.8}  & & 0.273\up{3.1}   & 0.793\up{3.9}   & 0.533\up{5.3} \\
\glove        & 0.253\up{4.0}   & 0.898\up{7.7}   & 0.626\up{5.8}   &    & 0.387\up{3.8}   & 0.940\up{5.6}   & 0.566\up{3.5}   & & 0.278\up{5.2}   & 0.799\up{4.7}   & 0.541\up{6.8}   \\
        \midrule
        \multicolumn{6}{l}{\bf \roberta{}} \\
        \baseline       & 0.243  & 0.812  & 0.557  &&  0.307  &  0.725  & 0.470 && 0.205   & 0.594   & 0.378 \\
\random       & 0.245\up{1.0}   & 0.878\up{4.1}   & 0.583\up{4.6}   &    & 0.365\up{18.8}   & 0.922\up{27.2}   & 0.557\up{18.5}$^{*}$   & & 0.257\up{25.8}   & 0.746\up{25.5}   & 0.496\up{37.4} \\
\bm           & 0.257\up{6.0}   & 0.873\up{7.5}   & 0.597\up{7.1}$^{*\#}$   &    & 0.362\up{18.1}    & 0.922\up{27.2}   & 0.548\up{16.7}$^{*}$ & & 0.265\up{29.7}   & 0.766\up{28.8}   & 0.509\up{34.9} \\
\glove        & 0.259\up{6.8}   & 0.863\up{6.3}   & 0.596\up{7.0}$^{*\#}$   &    & 0.354\up{15.3}   & 0.870\up{20.0}   & 0.550\up{17}$^{*}$  & & 0.267\up{30.4}   & 0.787\up{32.4}   & 0.519\up{37.3}  \\
        \midrule
        \multicolumn{6}{l}{\bf \distilbert{}} \\
        \baseline       & 0.244  & 0.843  & 0.565  &&  0.322  & 0.849   & 0.515 && 0.201   & 0.614   & 0.395 \\
        \random       & 0.253 \up{4}   & 0.883\up{4.7}   & 0.591\up{4.6}$^{\#}$   &    & 0.350\up{8.8}   & 0.919\up{8.2}   & 0.557\up{8.1}  & & 0.213\up{6.3}   & 0.713\up{16.2}   & 0.480\up{21.6}  \\
\bm           & 0.248\up{2.0}   & 0.909\up{7.8}   & 0.573\up{1.3}$^{\#}$   &    & 0.346\up{7.6}   & 0.915\up{7.7}   & 0.538\up{4.4}  & & 0.211\up{5.3}   & 0.704\up{14.7}   & 0.505\up{27.8}  \\
\glove        & 0.250\up{2.5}   & 0.872\up{3.8}   & 0.583\up{3.1}   &   & 0.338\up{5.1}   & 0.907\up{6.8}   & 0.505\down{1.9} & & 0.210\up{3.9}   & 0.681\up{11.0}   & 0.509\up{28.9}   \\
  
        \bottomrule
    \end{tabular}
    % } resize table end bracket
    \caption{Long document re-ranking results on the \trecdldf{}, \trecdlds{}, and \robust{} datasets. We train each ranker using three different data augmentation techniques on 10K instances. The models are trained using a pointwise variant of the \scl{} loss function. Statistically significant improvements at a level of $95\%$ and $90\%$ are indicated by $*$ and $\#$ respectively~\cite{paired_significance_test}.}
    \label{tab:performance_augmentation}
    %\vspace{-5mm}
\end{table*}

%% file: tables/table-bm25.tex
\begin{table*}
    \centering
    % \resizebox{\columnwidth}{!}{
    \begin{tabular}{lccccccc}
        \toprule
            %\textbf{Dataset Size}
            & \multicolumn{3}{c}{\ce{}} 
            && \multicolumn{3}{c}{\pairwise{}} \\
            \cmidrule(lr){2-4}
            \cmidrule(lr){6-8}
            \textit{Ranking Models}
            & $\text{AP}$ & $\text{RR}$ & $\text{nDCG}_\text{10}$ 
            && $\text{AP}$ & $\text{RR}$ & $\text{nDCG}_\text{10}$ \\
        \midrule
        \multicolumn{8}{l}{\bf \bert{}} \\
1k   	& 0.237\up{1.5} &	0.868\up{3.7} &	0.551\up{3.1}	&&	0.239\up{4.3}	 & 0.851\up{6.2}& 0.576\up{5.7}\\
2k    	& 0.241\up{1.9} &	0.916\up{12.9} & 0.592\up{5.2}	&&	0.248\up{3.6}	& 0.892\down{-0.4} & 0.603\up{1.5}	$^{*}$ \\
10k     & 0.258\up{5.8}&	0.924\up{10.8}&	0.617\up{4.3}	&&	0.264\up{3.1}& 	0.926\up{3.9}&	0.627\up{7.5}$^{*}$ \\
100k	& 0.260\up{1.8}&	0.942\up{4.3}	&0.653\up{6.3}	&&	0.270\up{0.6}&	0.959\up{2.7}&	0.666\up{3.4} \\
        \midrule
        \multicolumn{8}{l}{\bf \roberta{}} \\
        1k   & 0.170\up{2.3}&	0.697\up{25.9}&	0.319\up{7.4}	&&	0.228\up{25.9}&	0.803\up{15.7}&	0.533\up{59.8} \\
2k  & 0.171\up{1}&	0.670\up{12.4}&	0.322\up{9.5}	&&	0.236\up{4.4}	& 0.871\up{4.7}&	0.587\up{7.4} \\
10k & 0.257\up{6}&	0.873\up{7.5}&	0.597\up{7.1}$^{*\#}$	&&	0.261\up{3.5}&	0.914\up{3.8}&	0.633\up{3.5}$^{*}$ \\
100k & 0.263\up{2.9}&	0.946\up{4.7}&	0.646\up{11.7}	&&	0.270\up{1.2}& 	0.955\up{1.4}&	0.6667\up{0.3} \\
        \midrule
        \multicolumn{8}{l}{\bf \distilbert{}} \\
         1k   & 0.150\up{0}&	0.553\up{14.3}&	0.239\up{9.2}	&&	0.208\up{33.9}&	0.802\up{35.8}&	0.471\up{61.4}  \\
2k   & 0.164\up{2.3}&	0.589\up{0.6}	&0.304\up{9.2}	&&	0.231\up{15}&	0.862\up{13.1}&	0.526\up{19.4} \\
10k  & 0.248\up{2.0}&	0.909\up{7.8}	& 0.573\up{1.3}	$^{\#}$&&	0.253\up{5.1}	&0.893\up{3.9}&	0.613\up{7.7}$^{*}$    \\
100k & 0.255\up{1.1}&	0.942\up{3.1}&	0.641\up{5.7}	&&	0.270\up{3.3}	&0.927\up{2.9}&	0.645\up{1.5}$^{*}$ \\
        
        \bottomrule
    \end{tabular}
    % }
    \caption{Document re-ranking results on the \trecdldf{} datasets for \ce{} and \pairwise{} with \scl{} with data augmentation using BM25 selection strategy. We show the relative improvement of the augmentation approaches against a baseline without augmentation in parentheses. Statistically significant improvements at a level of $95\%$ and $90\%$ are indicated by $*$ and $\#$ respectively~\cite{paired_significance_test}.}
    \label{tab:pointwise-vs-pairwise}
    %\vspace{-5mm}
\end{table*}

%% file: figures.tex
\begin{figure*}
    \begin{subfigure}{.33\linewidth}
      \centering
      \includegraphics[width=\textwidth]{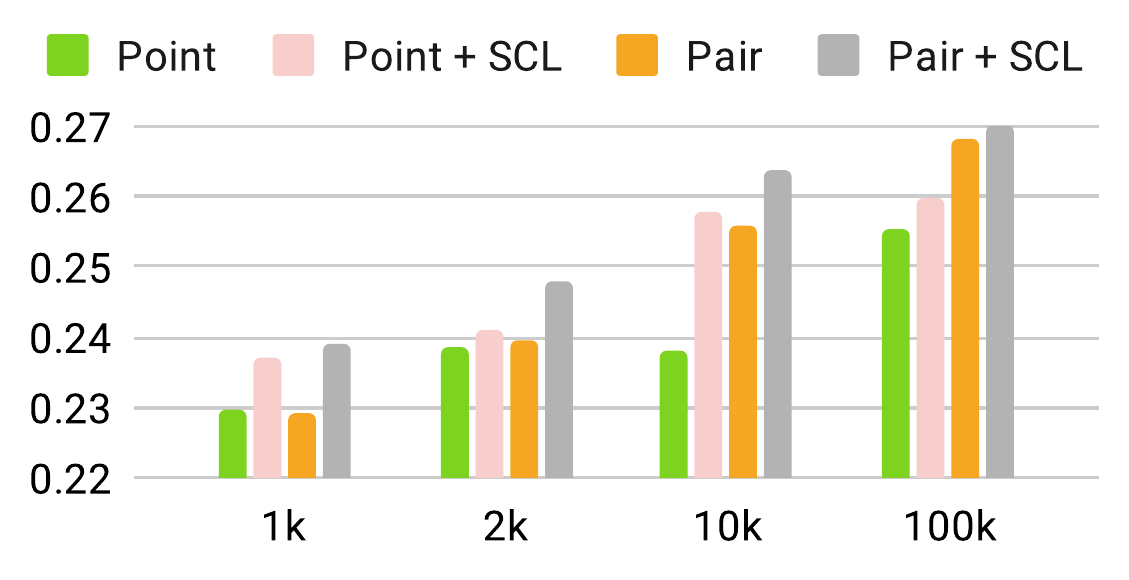}
      \caption{\bert{}}
    \end{subfigure}
    \begin{subfigure}{.33\linewidth}
      \centering
      \includegraphics[width=\textwidth]{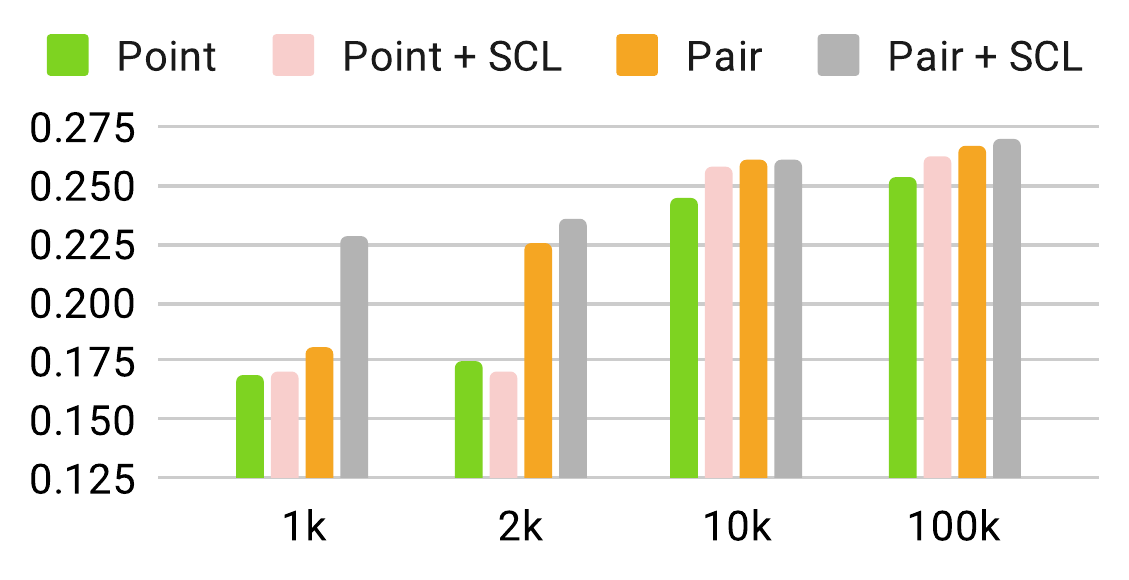}
      \caption{\roberta{}}
    \end{subfigure}
    \begin{subfigure}{.33\linewidth}
      \centering
      \includegraphics[width=\textwidth]{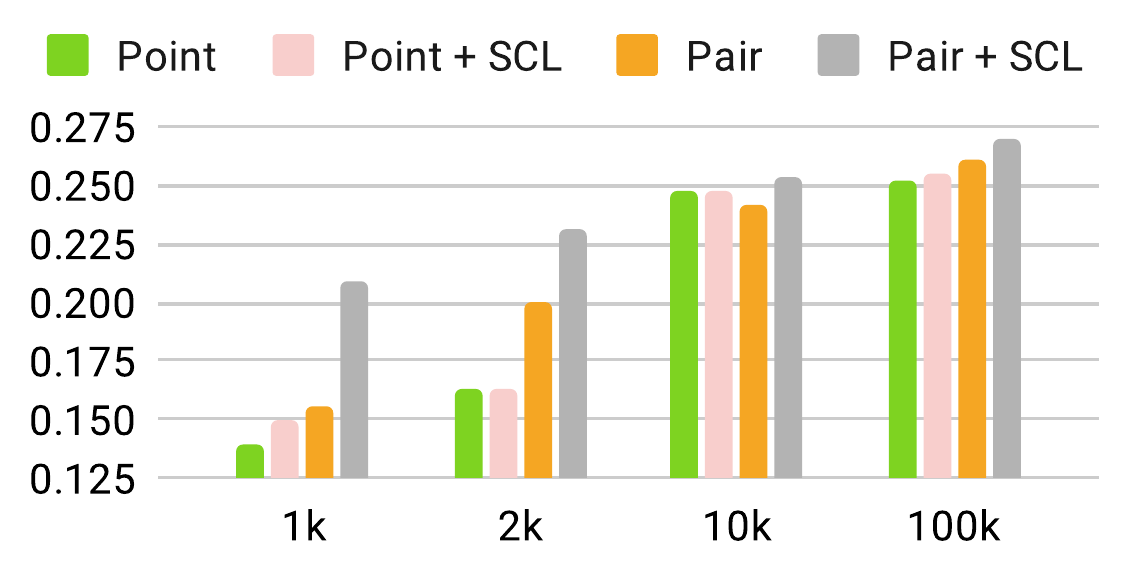}
      \caption{\distilbert{}}
    \end{subfigure}
    \caption{Comparing MAP score of \ce{} and \pairwise{} with corresponding \scl{} variant for different training instance sizes and models on \trecdl{}.}
    \label{fig:trainingset_size}
    %\vspace{-5mm}
\end{figure*}

%% file: tables/table-beir-robust-results.tex
      \begin{table*}
    \centering
    % \resizebox{\columnwidth}{!}{
    \begin{tabular}{lp{1.4cm}p{1.2cm}p{1.1cm}p{0.0cm}p{1.4cm}p{1.4cm}p{1.4cm}p{0.0cm}p{1.2cm}p{1.1cm}p{1.4cm}}
    %\begin{tabular}{cccccccccccc}
        \toprule
            & \multicolumn{3}{c}{\robust{}} 
            && \multicolumn{3}{c}{\scifact{}} 
             && \multicolumn{3}{c}{\fiqa{}}\\
            \cmidrule(lr){2-4}
            \cmidrule(lr){6-8}
            \cmidrule(lr){10-12}
            & $\text{AP}$ & $\text{RR}$ & $\text{nDCG}_\text{10}$ &
            & 
            $\text{AP}$ & $\text{RR}$ & $\text{nDCG}_\text{10}$ &
            & 
            $\text{AP}$ & $\text{RR}$ & $\text{nDCG}_\text{10}$\\
        \midrule
        \multicolumn{9}{l}{\bf \bert{}} \\
        \base{}-pointwise      & 0.264 &	0.763&	0.506   &  & 0.312 &	0.32 &	0.383& &0.140 &	0.221 &	0.187 \\
        \ce{}       & 0.276\up{4.7}   & 0.797\up{4.5}   & 0.537\up{6}   &  & 0.434\up{39}   & 0.448\up{40}   & 0.466\up{22}& &0.141\up{0.8}   & 0.221\up{3.4}   & 0.187\down{1.5} \\
        \base{}-pairwise & 0.195 &	0.599 &	0.382  &    & 0.454 &	0.466 &	0.504& &0.136 &	0.205 &	0.174 \\ 
\pairwise{} & 0.200\up{2.7}  & 0.601\up{0.4}  & 0.388\up{1.6}  &    & 0.562\up{33.6}  & 0.575\up{23.5}   & 0.616\up{29}$^{*}$& &0.221\up{63}  & 0.343\up{67}   & 0.277\up{59} \\ 
        \midrule
        \multicolumn{9}{l}{\bf \roberta{}} \\
        \base{}-pointwise       & 0.205 &	0.594 &	0.3776   &    & 0.615&	0.626	& 0.668& &0.113 &	0.173	& 0.146  \\
\ce{}       & 0.258\up{26}   & 0.746\up{25.5}   & 0.496\up{37.4}   &    & 0.638\up{3.7}   & 0.649\up{3.7}   & 0.687\up{2.8}$^{*}$& &0.240\up{112}   & 0.365\up{111}   & 0.300\up{108}  \\
\base{}-pairwise           & 0.250	& 0.762	&0.460	&&	0.641 &	0.652 &	0.685	&&	0.255 & 0.382 &	0.316\\
\pairwise{}           & 0.277\up{13.9}   & 0.529\up{11.65}   & 0.766\up{6.1}$^{*}$  &    & 0.668\up{4.2}   & 0.681\up{4.5}   & 0.712\up{3.8}$^{*}$& &0.274\up{7.6}   & 0.412\up{7.9}   & 0.339\up{7.4}$^{*}$\\
        \midrule
        \multicolumn{9}{l}{\bf \distilbert{}} \\
        \base{}-pointwise       & 0.201 &	0.614 &	0.395	&&	0.551	 & 0.567	 & 0.595	&&	0.111 &	0.188 &	0.132 \\
        \ce{}       & 0.258\up{28.5}   & 0.688\up{12.1}   & 0.480\up{21.6}   &    & 0.532\down{3.5}   & 0.558\down{3.3}   & 0.574\down{3.6}& &0.170\up{54}   & 0.269\up{43}   & 0.216\up{64}$^{*}$ \\
        \base{}-pairwise           & 0.186 &	0.372 &	0.576	&&	0.538	 & 0.554	 & 0.577	&&	0.235	& 0.362 &	0.288\\
\pairwise{}           & 0.182\down{1.9}   & 0.617\up{7}   & 0.375\up{0.7}$^{*}$   &    & 0.558\up{3.8}   & 0.573\up{3.4}   & 0.599\up{3.8}& &0.238\up{1.2}   & 0.366\up{1.2}   & 0.319\up{10.8}$^{*}$ \\
  
        \bottomrule
    \end{tabular}
    % } resize table end bracket
    \caption{Document re-ranking results on the \robust{}{}, \scifact{} and \fiqa{} datasets. We train each ranker using  \random{} data augmentation techniques on different datasets. The models are trained using a linear interpolation of \ce{} (\ce{} and \scl{}) and \pairwise{} (\pairwise{} and \scl{}) loss functions.  Values in brackets are percentage improvement from baseline. Statistically significant improvements at a level of $95\%$ and $90\%$ are indicated by $*$ and $\#$ respectively~\cite{paired_significance_test}.}
    \label{tab:robust_beir}
\end{table*}
%\vspace{-5mm}

%% file: conclusion.tex
\section{Discussion and Conclusion}
\label{sec:conclusion}
We make several important observations from our results.
Firstly, we find that just using augmented training data with existing \ce{} or \pairwise{} objectives does not result in performance improvements. 
In fact, in many scenarios the ranking performance decreases when using data augmentation with existing loss functions justifying existing work in the vision and language community that show the fragility of cross-entropy losses when using noisy labels~\cite{gunel2020supervised,khosla2020supervised}.
Instead we clearly show that \scl{} improves the ranking performance when using data augmentation in a variety of datasets.
Secondly, we find that this type of data augmentation surprisingly has little to no effect on the ranking performance. 
This suggests that using cheaper data augmentation schemes are already useful in simplifying the design decisions to be considered when using the \scl{} loss.
Finally, we observe that data augmentation is useful in improving the ranking performance specifically when training data sets are small and the marginal utility of data augmentation reduces with increasing data sizes with the maximum improvements being observed for low data settings.
We also observe that different inductive biases (contextual models) react differently to \scl{}, with \roberta{}-based ranker showing improvements in ranking metrics up to $ > 50\%$ for smaller datasets over its non-augmented counterparts. 

%% file: main.bbl

\begin{thebibliography}{79}

%%% the Web address.

\ifx \showCODEN    \undefined \def \showCODEN     #1{\unskip}     \fi
\ifx \showDOI      \undefined \def \showDOI       #1{#1}\fi
\ifx \showISBNx    \undefined \def \showISBNx     #1{\unskip}     \fi
\ifx \showISBNxiii \undefined \def \showISBNxiii  #1{\unskip}     \fi
\ifx \showISSN     \undefined \def \showISSN      #1{\unskip}     \fi
\ifx \showLCCN     \undefined \def \showLCCN      #1{\unskip}     \fi
\ifx \shownote     \undefined \def \shownote      #1{#1}          \fi
\ifx \showarticletitle \undefined \def \showarticletitle #1{#1}   \fi
\ifx \showURL      \undefined \def \showURL       {\relax}        \fi
% invisible to TeX
\providecommand\bibfield[2]{#2}
\providecommand\bibinfo[2]{#2}
\providecommand\natexlab[1]{#1}
\providecommand\showeprint[2][]{arXiv:#2}

\bibitem[\protect\citeauthoryear{Ahmad, Constant, Yang, and Cer}{Ahmad
  et~al\mbox{.}}{2019}]%
        {ahmad2019reqa}
\bibfield{author}{\bibinfo{person}{Amin Ahmad}, \bibinfo{person}{Noah
  Constant}, \bibinfo{person}{Yinfei Yang}, {and} \bibinfo{person}{Daniel
  Cer}.} \bibinfo{year}{2019}\natexlab{}.
\newblock \showarticletitle{ReQA: An evaluation for end-to-end answer retrieval
  models}.
\newblock \bibinfo{journal}{\emph{arXiv preprint arXiv:1907.04780}}
  (\bibinfo{year}{2019}).
\newblock


\bibitem[\protect\citeauthoryear{Chang, Yu, Chang, Yang, and Kumar}{Chang
  et~al\mbox{.}}{2020}]%
        {chang2020pre}
\bibfield{author}{\bibinfo{person}{Wei-Cheng Chang}, \bibinfo{person}{Felix~X
  Yu}, \bibinfo{person}{Yin-Wen Chang}, \bibinfo{person}{Yiming Yang}, {and}
  \bibinfo{person}{Sanjiv Kumar}.} \bibinfo{year}{2020}\natexlab{}.
\newblock \showarticletitle{Pre-training tasks for embedding-based large-scale
  retrieval}.
\newblock \bibinfo{journal}{\emph{arXiv preprint arXiv:2002.03932}}
  (\bibinfo{year}{2020}).
\newblock


\bibitem[\protect\citeauthoryear{Chen, Liu, Zhao, and Jia}{Chen
  et~al\mbox{.}}{2020}]%
        {chen_2020_gridmask}
\bibfield{author}{\bibinfo{person}{Pengguang Chen}, \bibinfo{person}{Shu Liu},
  \bibinfo{person}{Hengshuang Zhao}, {and} \bibinfo{person}{Jiaya Jia}.}
  \bibinfo{year}{2020}\natexlab{}.
\newblock \showarticletitle{Gridmask data augmentation}.
\newblock \bibinfo{journal}{\emph{arXiv preprint arXiv:2001.04086}}
  (\bibinfo{year}{2020}).
\newblock


\bibitem[\protect\citeauthoryear{Chen, Kedzie, Nair,
  Galu{\v{s}}{\v{c}}{\'a}kov{\'a}, Zhang, Oard, and McKeown}{Chen
  et~al\mbox{.}}{2021}]%
        {chen_2021_cross}
\bibfield{author}{\bibinfo{person}{Yanda Chen}, \bibinfo{person}{Chris Kedzie},
  \bibinfo{person}{Suraj Nair}, \bibinfo{person}{Petra
  Galu{\v{s}}{\v{c}}{\'a}kov{\'a}}, \bibinfo{person}{Rui Zhang},
  \bibinfo{person}{Douglas~W Oard}, {and} \bibinfo{person}{Kathleen McKeown}.}
  \bibinfo{year}{2021}\natexlab{}.
\newblock \showarticletitle{Cross-language Sentence Selection via Data
  Augmentation and Rationale Training}.
\newblock \bibinfo{journal}{\emph{arXiv preprint arXiv:2106.02293}}
  (\bibinfo{year}{2021}).
\newblock


\bibitem[\protect\citeauthoryear{Craswell, Mitra, Yilmaz, and Campos}{Craswell
  et~al\mbox{.}}{2019}]%
        {msmarco_trec_2019}
\bibfield{author}{\bibinfo{person}{Nick Craswell}, \bibinfo{person}{Bhaskar
  Mitra}, \bibinfo{person}{Emine Yilmaz}, {and} \bibinfo{person}{Daniel
  Campos}.} \bibinfo{year}{2019}\natexlab{}.
\newblock \bibinfo{title}{{TREC-2019-Deep-Learning}}.
\newblock
  \bibinfo{howpublished}{\url{https://microsoft.github.io/TREC-2019-Deep-Learning/}}.
\newblock


\bibitem[\protect\citeauthoryear{Cubuk, Zoph, Mane, Vasudevan, and Le}{Cubuk
  et~al\mbox{.}}{2019}]%
        {cubuk_2019_autoaugment}
\bibfield{author}{\bibinfo{person}{Ekin~D Cubuk}, \bibinfo{person}{Barret
  Zoph}, \bibinfo{person}{Dandelion Mane}, \bibinfo{person}{Vijay Vasudevan},
  {and} \bibinfo{person}{Quoc~V Le}.} \bibinfo{year}{2019}\natexlab{}.
\newblock \showarticletitle{Autoaugment: Learning augmentation strategies from
  data}. In \bibinfo{booktitle}{\emph{Proceedings of the IEEE/CVF Conference on
  Computer Vision and Pattern Recognition}}. \bibinfo{pages}{113--123}.
\newblock


\bibitem[\protect\citeauthoryear{Dai and Callan}{Dai and Callan}{2019}]%
        {dai_sigir_2019}
\bibfield{author}{\bibinfo{person}{Zhuyun Dai} {and} \bibinfo{person}{Jamie
  Callan}.} \bibinfo{year}{2019}\natexlab{}.
\newblock \showarticletitle{Deeper Text Understanding for IR with Contextual
  Neural Language Modeling}. In \bibinfo{booktitle}{\emph{ACM SIGIR'19}}.
  \bibinfo{pages}{985--988}.
\newblock


\bibitem[\protect\citeauthoryear{Devlin, Chang, Lee, and Toutanova}{Devlin
  et~al\mbox{.}}{2018}]%
        {devlin_bert_2018}
\bibfield{author}{\bibinfo{person}{Jacob Devlin}, \bibinfo{person}{Ming{-}Wei
  Chang}, \bibinfo{person}{Kenton Lee}, {and} \bibinfo{person}{Kristina
  Toutanova}.} \bibinfo{year}{2018}\natexlab{}.
\newblock \showarticletitle{{BERT:} Pre-training of Deep Bidirectional
  Transformers for Language Understanding}.
\newblock \bibinfo{journal}{\emph{CoRR}}  \bibinfo{volume}{abs/1810.04805}
  (\bibinfo{year}{2018}).
\newblock
\urldef\tempurl%
\url{http://arxiv.org/abs/1810.04805}
\showURL{%
\tempurl}


\bibitem[\protect\citeauthoryear{Dodge, Ilharco, Schwartz, Farhadi, Hajishirzi,
  and Smith}{Dodge et~al\mbox{.}}{2020}]%
        {dodge2020fine}
\bibfield{author}{\bibinfo{person}{Jesse Dodge}, \bibinfo{person}{Gabriel
  Ilharco}, \bibinfo{person}{Roy Schwartz}, \bibinfo{person}{Ali Farhadi},
  \bibinfo{person}{Hannaneh Hajishirzi}, {and} \bibinfo{person}{Noah Smith}.}
  \bibinfo{year}{2020}\natexlab{}.
\newblock \showarticletitle{Fine-tuning pretrained language models: Weight
  initializations, data orders, and early stopping}.
\newblock \bibinfo{journal}{\emph{arXiv preprint arXiv:2002.06305}}
  (\bibinfo{year}{2020}).
\newblock


\bibitem[\protect\citeauthoryear{Donahue and Simonyan}{Donahue and
  Simonyan}{2019}]%
        {donahue2019large}
\bibfield{author}{\bibinfo{person}{Jeff Donahue} {and} \bibinfo{person}{Karen
  Simonyan}.} \bibinfo{year}{2019}\natexlab{}.
\newblock \showarticletitle{Large scale adversarial representation learning}.
\newblock \bibinfo{journal}{\emph{arXiv preprint arXiv:1907.02544}}
  (\bibinfo{year}{2019}).
\newblock


\bibitem[\protect\citeauthoryear{Dong, Zhu, Fu, Xu, and de~Melo}{Dong
  et~al\mbox{.}}{2021}]%
        {dong_2021_cross_nli}
\bibfield{author}{\bibinfo{person}{Xin~Luna Dong}, \bibinfo{person}{Yaxin Zhu},
  \bibinfo{person}{Zuohui Fu}, \bibinfo{person}{Dongkuan Xu}, {and}
  \bibinfo{person}{Gerard de Melo}.} \bibinfo{year}{2021}\natexlab{}.
\newblock \showarticletitle{Data Augmentation with Adversarial Training for
  Cross-Lingual NLI}. In \bibinfo{booktitle}{\emph{Proceedings of the 59th
  Annual Meeting of the Association for Computational Linguistics and the 11th
  International Joint Conference on Natural Language Processing (Volume 1: Long
  Papers)}}. \bibinfo{pages}{5158--5167}.
\newblock


\bibitem[\protect\citeauthoryear{Gallagher}{Gallagher}{2019}]%
        {paired_significance_test}
\bibfield{author}{\bibinfo{person}{Luke Gallagher}.}
  \bibinfo{year}{2019}\natexlab{}.
\newblock \bibinfo{title}{{Pairwise t-test on TREC Run Files}}.
\newblock
  \bibinfo{howpublished}{\url{https://github.com/lgrz/pairwise-ttest/}}.
\newblock


\bibitem[\protect\citeauthoryear{Gao, Saha, Prasad, and Roychoudhury}{Gao
  et~al\mbox{.}}{2020}]%
        {gao_2020_fuzz}
\bibfield{author}{\bibinfo{person}{Xiang Gao}, \bibinfo{person}{Ripon~K Saha},
  \bibinfo{person}{Mukul~R Prasad}, {and} \bibinfo{person}{Abhik
  Roychoudhury}.} \bibinfo{year}{2020}\natexlab{}.
\newblock \showarticletitle{Fuzz testing based data augmentation to improve
  robustness of deep neural networks}. In \bibinfo{booktitle}{\emph{2020
  IEEE/ACM 42nd International Conference on Software Engineering (ICSE)}}.
  IEEE, \bibinfo{pages}{1147--1158}.
\newblock


\bibitem[\protect\citeauthoryear{Gidaris, Singh, and Komodakis}{Gidaris
  et~al\mbox{.}}{2018}]%
        {gidaris2018unsupervised}
\bibfield{author}{\bibinfo{person}{Spyros Gidaris}, \bibinfo{person}{Praveer
  Singh}, {and} \bibinfo{person}{Nikos Komodakis}.}
  \bibinfo{year}{2018}\natexlab{}.
\newblock \showarticletitle{Unsupervised representation learning by predicting
  image rotations}.
\newblock \bibinfo{journal}{\emph{arXiv preprint arXiv:1803.07728}}
  (\bibinfo{year}{2018}).
\newblock


\bibitem[\protect\citeauthoryear{Gunel, Du, Conneau, and Stoyanov}{Gunel
  et~al\mbox{.}}{2020}]%
        {gunel2020supervised}
\bibfield{author}{\bibinfo{person}{Beliz Gunel}, \bibinfo{person}{Jingfei Du},
  \bibinfo{person}{Alexis Conneau}, {and} \bibinfo{person}{Ves Stoyanov}.}
  \bibinfo{year}{2020}\natexlab{}.
\newblock \showarticletitle{Supervised contrastive learning for pre-trained
  language model fine-tuning}.
\newblock \bibinfo{journal}{\emph{arXiv preprint arXiv:2011.01403}}
  (\bibinfo{year}{2020}).
\newblock


\bibitem[\protect\citeauthoryear{He, Fan, Wu, Xie, and Girshick}{He
  et~al\mbox{.}}{2020}]%
        {he2020momentum}
\bibfield{author}{\bibinfo{person}{Kaiming He}, \bibinfo{person}{Haoqi Fan},
  \bibinfo{person}{Yuxin Wu}, \bibinfo{person}{Saining Xie}, {and}
  \bibinfo{person}{Ross Girshick}.} \bibinfo{year}{2020}\natexlab{}.
\newblock \showarticletitle{Momentum contrast for unsupervised visual
  representation learning}. In \bibinfo{booktitle}{\emph{Proceedings of the
  IEEE/CVF Conference on Computer Vision and Pattern Recognition}}.
  \bibinfo{pages}{9729--9738}.
\newblock


\bibitem[\protect\citeauthoryear{Hjelm, Fedorov, Lavoie-Marchildon, Grewal,
  Bachman, Trischler, and Bengio}{Hjelm et~al\mbox{.}}{2018}]%
        {hjelm2018learning}
\bibfield{author}{\bibinfo{person}{R~Devon Hjelm}, \bibinfo{person}{Alex
  Fedorov}, \bibinfo{person}{Samuel Lavoie-Marchildon}, \bibinfo{person}{Karan
  Grewal}, \bibinfo{person}{Phil Bachman}, \bibinfo{person}{Adam Trischler},
  {and} \bibinfo{person}{Yoshua Bengio}.} \bibinfo{year}{2018}\natexlab{}.
\newblock \showarticletitle{Learning deep representations by mutual information
  estimation and maximization}.
\newblock \bibinfo{journal}{\emph{arXiv preprint arXiv:1808.06670}}
  (\bibinfo{year}{2018}).
\newblock


\bibitem[\protect\citeauthoryear{Hofst{\"a}tter, Zamani, Mitra, Craswell, and
  Hanbury}{Hofst{\"a}tter et~al\mbox{.}}{2020a}]%
        {hofstatter2020local}
\bibfield{author}{\bibinfo{person}{Sebastian Hofst{\"a}tter},
  \bibinfo{person}{Hamed Zamani}, \bibinfo{person}{Bhaskar Mitra},
  \bibinfo{person}{Nick Craswell}, {and} \bibinfo{person}{Allan Hanbury}.}
  \bibinfo{year}{2020}\natexlab{a}.
\newblock \showarticletitle{Local Self-Attention over Long Text for Efficient
  Document Retrieval}.
\newblock \bibinfo{journal}{\emph{arXiv preprint arXiv:2005.04908}}
  (\bibinfo{year}{2020}).
\newblock


\bibitem[\protect\citeauthoryear{Hofst{\"a}tter, Zlabinger, and
  Hanbury}{Hofst{\"a}tter et~al\mbox{.}}{2020b}]%
        {hofstatter2020interpretable}
\bibfield{author}{\bibinfo{person}{Sebastian Hofst{\"a}tter},
  \bibinfo{person}{Markus Zlabinger}, {and} \bibinfo{person}{Allan Hanbury}.}
  \bibinfo{year}{2020}\natexlab{b}.
\newblock \showarticletitle{Interpretable \& time-budget-constrained
  contextualization for re-ranking}.
\newblock \bibinfo{journal}{\emph{arXiv preprint arXiv:2002.01854}}
  (\bibinfo{year}{2020}).
\newblock


\bibitem[\protect\citeauthoryear{Jaiswal, Babu, Zadeh, Banerjee, and
  Makedon}{Jaiswal et~al\mbox{.}}{2021}]%
        {jaiswal2021survey}
\bibfield{author}{\bibinfo{person}{Ashish Jaiswal},
  \bibinfo{person}{Ashwin~Ramesh Babu}, \bibinfo{person}{Mohammad~Zaki Zadeh},
  \bibinfo{person}{Debapriya Banerjee}, {and} \bibinfo{person}{Fillia
  Makedon}.} \bibinfo{year}{2021}\natexlab{}.
\newblock \showarticletitle{A survey on contrastive self-supervised learning}.
\newblock \bibinfo{journal}{\emph{Technologies}} \bibinfo{volume}{9},
  \bibinfo{number}{1} (\bibinfo{year}{2021}), \bibinfo{pages}{2}.
\newblock


\bibitem[\protect\citeauthoryear{Karpukhin, O{\u{g}}uz, Min, Lewis, Wu, Edunov,
  Chen, and Yih}{Karpukhin et~al\mbox{.}}{2020}]%
        {karpukhin2020dense}
\bibfield{author}{\bibinfo{person}{Vladimir Karpukhin}, \bibinfo{person}{Barlas
  O{\u{g}}uz}, \bibinfo{person}{Sewon Min}, \bibinfo{person}{Patrick Lewis},
  \bibinfo{person}{Ledell Wu}, \bibinfo{person}{Sergey Edunov},
  \bibinfo{person}{Danqi Chen}, {and} \bibinfo{person}{Wen-tau Yih}.}
  \bibinfo{year}{2020}\natexlab{}.
\newblock \showarticletitle{Dense passage retrieval for open-domain question
  answering}.
\newblock \bibinfo{journal}{\emph{arXiv preprint arXiv:2004.04906}}
  (\bibinfo{year}{2020}).
\newblock


\bibitem[\protect\citeauthoryear{Khattab and Zaharia}{Khattab and
  Zaharia}{2020}]%
        {khattab2020colbert}
\bibfield{author}{\bibinfo{person}{Omar Khattab} {and} \bibinfo{person}{Matei
  Zaharia}.} \bibinfo{year}{2020}\natexlab{}.
\newblock \showarticletitle{ColBERT: Efficient and Effective Passage Search via
  Contextualized Late Interaction over BERT}.
\newblock
\showeprint[arxiv]{2004.12832}


\bibitem[\protect\citeauthoryear{Khosla, Teterwak, Wang, Sarna, Tian, Isola,
  Maschinot, Liu, and Krishnan}{Khosla et~al\mbox{.}}{2020}]%
        {khosla2020supervised}
\bibfield{author}{\bibinfo{person}{Prannay Khosla}, \bibinfo{person}{Piotr
  Teterwak}, \bibinfo{person}{Chen Wang}, \bibinfo{person}{Aaron Sarna},
  \bibinfo{person}{Yonglong Tian}, \bibinfo{person}{Phillip Isola},
  \bibinfo{person}{Aaron Maschinot}, \bibinfo{person}{Ce Liu}, {and}
  \bibinfo{person}{Dilip Krishnan}.} \bibinfo{year}{2020}\natexlab{}.
\newblock \showarticletitle{Supervised contrastive learning}.
\newblock \bibinfo{journal}{\emph{Advances in Neural Information Processing
  Systems}}  \bibinfo{volume}{33} (\bibinfo{year}{2020}),
  \bibinfo{pages}{18661--18673}.
\newblock


\bibitem[\protect\citeauthoryear{Kumar, Choudhary, and Cho}{Kumar
  et~al\mbox{.}}{2020}]%
        {kumar_2020_augment_transformer}
\bibfield{author}{\bibinfo{person}{Varun Kumar}, \bibinfo{person}{Ashutosh
  Choudhary}, {and} \bibinfo{person}{Eunah Cho}.}
  \bibinfo{year}{2020}\natexlab{}.
\newblock \showarticletitle{Data augmentation using pre-trained transformer
  models}.
\newblock \bibinfo{journal}{\emph{arXiv preprint arXiv:2003.02245}}
  (\bibinfo{year}{2020}).
\newblock


\bibitem[\protect\citeauthoryear{Lee, Chang, and Toutanova}{Lee
  et~al\mbox{.}}{2019}]%
        {lee2019latent}
\bibfield{author}{\bibinfo{person}{Kenton Lee}, \bibinfo{person}{Ming-Wei
  Chang}, {and} \bibinfo{person}{Kristina Toutanova}.}
  \bibinfo{year}{2019}\natexlab{}.
\newblock \showarticletitle{Latent retrieval for weakly supervised open domain
  question answering}.
\newblock \bibinfo{journal}{\emph{arXiv preprint arXiv:1906.00300}}
  (\bibinfo{year}{2019}).
\newblock


\bibitem[\protect\citeauthoryear{Leonhardt, Rudra, and Anand}{Leonhardt
  et~al\mbox{.}}{2021a}]%
        {leonhardt2021learnt}
\bibfield{author}{\bibinfo{person}{Jurek Leonhardt}, \bibinfo{person}{Koustav
  Rudra}, {and} \bibinfo{person}{Avishek Anand}.}
  \bibinfo{year}{2021}\natexlab{a}.
\newblock \showarticletitle{Learnt Sparsity for Effective and Interpretable
  Document Ranking}.
\newblock \bibinfo{journal}{\emph{arXiv preprint arXiv:2106.12460}}
  (\bibinfo{year}{2021}).
\newblock


\bibitem[\protect\citeauthoryear{Leonhardt, Rudra, Khosla, Anand, and
  Anand}{Leonhardt et~al\mbox{.}}{2021b}]%
        {leonhardt2021fast}
\bibfield{author}{\bibinfo{person}{Jurek Leonhardt}, \bibinfo{person}{Koustav
  Rudra}, \bibinfo{person}{Megha Khosla}, \bibinfo{person}{Abhijit Anand},
  {and} \bibinfo{person}{Avishek Anand}.} \bibinfo{year}{2021}\natexlab{b}.
\newblock \showarticletitle{Fast Forward Indexes for Efficient Document
  Ranking}.
\newblock \bibinfo{journal}{\emph{arXiv preprint arXiv:2110.06051}}
  (\bibinfo{year}{2021}).
\newblock


\bibitem[\protect\citeauthoryear{Li, Yates, MacAvaney, He, and Sun}{Li
  et~al\mbox{.}}{2020a}]%
        {li2020parade}
\bibfield{author}{\bibinfo{person}{Canjia Li}, \bibinfo{person}{Andrew Yates},
  \bibinfo{person}{Sean MacAvaney}, \bibinfo{person}{Ben He}, {and}
  \bibinfo{person}{Yingfei Sun}.} \bibinfo{year}{2020}\natexlab{a}.
\newblock \showarticletitle{PARADE: Passage Representation Aggregation for
  Document Reranking}.
\newblock \bibinfo{journal}{\emph{arXiv preprint arXiv:2008.09093}}
  (\bibinfo{year}{2020}).
\newblock


\bibitem[\protect\citeauthoryear{Li, Zhang, Tian, and Xiong}{Li
  et~al\mbox{.}}{2020b}]%
        {li_2020_recognition}
\bibfield{author}{\bibinfo{person}{Hao Li}, \bibinfo{person}{Xiaopeng Zhang},
  \bibinfo{person}{Qi Tian}, {and} \bibinfo{person}{Hongkai Xiong}.}
  \bibinfo{year}{2020}\natexlab{b}.
\newblock \showarticletitle{Attribute mix: Semantic data augmentation for fine
  grained recognition}. In \bibinfo{booktitle}{\emph{2020 IEEE International
  Conference on Visual Communications and Image Processing (VCIP)}}. IEEE,
  \bibinfo{pages}{243--246}.
\newblock


\bibitem[\protect\citeauthoryear{Li, Liu, Xiong, and Liu}{Li
  et~al\mbox{.}}{2021a}]%
        {li_2021_dual_learning}
\bibfield{author}{\bibinfo{person}{Yizhi Li}, \bibinfo{person}{Zhenghao Liu},
  \bibinfo{person}{Chenyan Xiong}, {and} \bibinfo{person}{Zhiyuan Liu}.}
  \bibinfo{year}{2021}\natexlab{a}.
\newblock \showarticletitle{More Robust Dense Retrieval with Contrastive Dual
  Learning}. In \bibinfo{booktitle}{\emph{Proceedings of the 2021 ACM SIGIR
  International Conference on Theory of Information Retrieval}}.
  \bibinfo{pages}{287–296}.
\newblock


\bibitem[\protect\citeauthoryear{Li, Liu, Xiong, and Liu}{Li
  et~al\mbox{.}}{2021b}]%
        {li2021more}
\bibfield{author}{\bibinfo{person}{Yizhi Li}, \bibinfo{person}{Zhenghao Liu},
  \bibinfo{person}{Chenyan Xiong}, {and} \bibinfo{person}{Zhiyuan Liu}.}
  \bibinfo{year}{2021}\natexlab{b}.
\newblock \showarticletitle{More Robust Dense Retrieval with Contrastive Dual
  Learning}. In \bibinfo{booktitle}{\emph{Proceedings of the 2021 ACM SIGIR
  International Conference on Theory of Information Retrieval}}.
  \bibinfo{pages}{287--296}.
\newblock


\bibitem[\protect\citeauthoryear{Lian, You, Wu, Liu, and Jia}{Lian
  et~al\mbox{.}}{2020}]%
        {lian_2020_retrieve_keyword}
\bibfield{author}{\bibinfo{person}{Yijiang Lian}, \bibinfo{person}{Zhenjun
  You}, \bibinfo{person}{Fan Wu}, \bibinfo{person}{Wenqiang Liu}, {and}
  \bibinfo{person}{Jing Jia}.} \bibinfo{year}{2020}\natexlab{}.
\newblock \showarticletitle{Retrieve Synonymous keywords for Frequent Queries
  in Sponsored Search in a Data Augmentation Way}.
\newblock \bibinfo{journal}{\emph{arXiv preprint arXiv:2008.01969}}
  (\bibinfo{year}{2020}).
\newblock


\bibitem[\protect\citeauthoryear{Lin}{Lin}{2019}]%
        {lin2019neural}
\bibfield{author}{\bibinfo{person}{Jimmy Lin}.}
  \bibinfo{year}{2019}\natexlab{}.
\newblock \showarticletitle{The Neural Hype and Comparisons Against Weak
  Baselines}. In \bibinfo{booktitle}{\emph{ACM SIGIR Forum}},
  Vol.~\bibinfo{volume}{52}. ACM, \bibinfo{pages}{40--51}.
\newblock


\bibitem[\protect\citeauthoryear{Liu, Ding, Bing, Joty, Si, and Miao}{Liu
  et~al\mbox{.}}{2021}]%
        {liu_mulda_low_resource_aug}
\bibfield{author}{\bibinfo{person}{Linlin Liu}, \bibinfo{person}{Bosheng Ding},
  \bibinfo{person}{Lidong Bing}, \bibinfo{person}{Shafiq~R. Joty},
  \bibinfo{person}{Luo Si}, {and} \bibinfo{person}{Chunyan Miao}.}
  \bibinfo{year}{2021}\natexlab{}.
\newblock \showarticletitle{MulDA: A Multilingual Data Augmentation Framework
  for Low-Resource Cross-Lingual NER}. In
  \bibinfo{booktitle}{\emph{ACL/IJCNLP}}. \bibinfo{pages}{5834--5846}.
\newblock
\urldef\tempurl%
\url{https://doi.org/10.18653/v1/2021.acl-long.453}
\showURL{%
\tempurl}


\bibitem[\protect\citeauthoryear{Liu, Wen, Yu, and Yang}{Liu
  et~al\mbox{.}}{2016}]%
        {liu2016large:augprob}
\bibfield{author}{\bibinfo{person}{Weiyang Liu}, \bibinfo{person}{Yandong Wen},
  \bibinfo{person}{Zhiding Yu}, {and} \bibinfo{person}{Meng Yang}.}
  \bibinfo{year}{2016}\natexlab{}.
\newblock \showarticletitle{Large-margin softmax loss for convolutional neural
  networks.}. In \bibinfo{booktitle}{\emph{ICML}}, Vol.~\bibinfo{volume}{2}.
  \bibinfo{pages}{7}.
\newblock


\bibitem[\protect\citeauthoryear{Liu, Ott, Goyal, Du, Joshi, Chen, Levy, Lewis,
  Zettlemoyer, and Stoyanov}{Liu et~al\mbox{.}}{2019}]%
        {liu2019roberta}
\bibfield{author}{\bibinfo{person}{Yinhan Liu}, \bibinfo{person}{Myle Ott},
  \bibinfo{person}{Naman Goyal}, \bibinfo{person}{Jingfei Du},
  \bibinfo{person}{Mandar Joshi}, \bibinfo{person}{Danqi Chen},
  \bibinfo{person}{Omer Levy}, \bibinfo{person}{Mike Lewis},
  \bibinfo{person}{Luke Zettlemoyer}, {and} \bibinfo{person}{Veselin
  Stoyanov}.} \bibinfo{year}{2019}\natexlab{}.
\newblock \showarticletitle{Roberta: A robustly optimized bert pretraining
  approach}.
\newblock \bibinfo{journal}{\emph{arXiv preprint arXiv:1907.11692}}
  (\bibinfo{year}{2019}).
\newblock


\bibitem[\protect\citeauthoryear{Longpre, Wang, and DuBois}{Longpre
  et~al\mbox{.}}{2020}]%
        {longpre_2020_effective}
\bibfield{author}{\bibinfo{person}{Shayne Longpre}, \bibinfo{person}{Yu Wang},
  {and} \bibinfo{person}{Christopher DuBois}.} \bibinfo{year}{2020}\natexlab{}.
\newblock \showarticletitle{How Effective is Task-Agnostic Data Augmentation
  for Pretrained Transformers?}
\newblock \bibinfo{journal}{\emph{arXiv preprint arXiv:2010.01764}}
  (\bibinfo{year}{2020}).
\newblock


\bibitem[\protect\citeauthoryear{Lu, Zheng, Velivelli, and Zhai}{Lu
  et~al\mbox{.}}{2006}]%
        {lu_2006_enhancing}
\bibfield{author}{\bibinfo{person}{Xinghua Lu}, \bibinfo{person}{Bin Zheng},
  \bibinfo{person}{Atulya Velivelli}, {and} \bibinfo{person}{ChengXiang Zhai}.}
  \bibinfo{year}{2006}\natexlab{}.
\newblock \showarticletitle{Enhancing text categorization with
  semantic-enriched representation and training data augmentation}.
\newblock \bibinfo{journal}{\emph{Journal of the American Medical Informatics
  Association}} \bibinfo{volume}{13}, \bibinfo{number}{5}
  (\bibinfo{year}{2006}), \bibinfo{pages}{526--535}.
\newblock


\bibitem[\protect\citeauthoryear{Luo, Zhu, Jin, and Wang}{Luo
  et~al\mbox{.}}{2020}]%
        {luo_2020_learn_augment}
\bibfield{author}{\bibinfo{person}{Canjie Luo}, \bibinfo{person}{Yuanzhi Zhu},
  \bibinfo{person}{Lianwen Jin}, {and} \bibinfo{person}{Yongpan Wang}.}
  \bibinfo{year}{2020}\natexlab{}.
\newblock \showarticletitle{Learn to augment: Joint data augmentation and
  network optimization for text recognition}. In
  \bibinfo{booktitle}{\emph{Proceedings of the IEEE/CVF Conference on Computer
  Vision and Pattern Recognition}}. \bibinfo{pages}{13746--13755}.
\newblock


\bibitem[\protect\citeauthoryear{MacAvaney, Yates, Cohan, and
  Goharian}{MacAvaney et~al\mbox{.}}{2019}]%
        {macavaney2019contextualized:bertir:mpi}
\bibfield{author}{\bibinfo{person}{Sean MacAvaney}, \bibinfo{person}{Andrew
  Yates}, \bibinfo{person}{Arman Cohan}, {and} \bibinfo{person}{Nazli
  Goharian}.} \bibinfo{year}{2019}\natexlab{}.
\newblock \showarticletitle{Contextualized Word Representations for Document
  Re-Ranking}.
\newblock \bibinfo{journal}{\emph{arXiv preprint arXiv:1904.07094}}
  (\bibinfo{year}{2019}).
\newblock


\bibitem[\protect\citeauthoryear{Mekruksavanich and
  Jitpattanakul}{Mekruksavanich and Jitpattanakul}{2021}]%
        {mekruksavanich_2021_user_identify}
\bibfield{author}{\bibinfo{person}{Sakorn Mekruksavanich} {and}
  \bibinfo{person}{Anuchit Jitpattanakul}.} \bibinfo{year}{2021}\natexlab{}.
\newblock \showarticletitle{Convolutional neural network and data augmentation
  for behavioral-based biometric user identification}.
\newblock In \bibinfo{booktitle}{\emph{ICT Systems and Sustainability}}.
  \bibinfo{publisher}{Springer}, \bibinfo{pages}{753--761}.
\newblock


\bibitem[\protect\citeauthoryear{Morris, Lifland, Yoo, Grigsby, Jin, and
  Qi}{Morris et~al\mbox{.}}{2020}]%
        {morris_2020_textattack}
\bibfield{author}{\bibinfo{person}{John~X Morris}, \bibinfo{person}{Eli
  Lifland}, \bibinfo{person}{Jin~Yong Yoo}, \bibinfo{person}{Jake Grigsby},
  \bibinfo{person}{Di Jin}, {and} \bibinfo{person}{Yanjun Qi}.}
  \bibinfo{year}{2020}\natexlab{}.
\newblock \showarticletitle{Textattack: A framework for adversarial attacks,
  data augmentation, and adversarial training in nlp}.
\newblock \bibinfo{journal}{\emph{arXiv preprint arXiv:2005.05909}}
  (\bibinfo{year}{2020}).
\newblock


\bibitem[\protect\citeauthoryear{Mosbach, Andriushchenko, and Klakow}{Mosbach
  et~al\mbox{.}}{2020}]%
        {mosbach2020stability}
\bibfield{author}{\bibinfo{person}{Marius Mosbach}, \bibinfo{person}{Maksym
  Andriushchenko}, {and} \bibinfo{person}{Dietrich Klakow}.}
  \bibinfo{year}{2020}\natexlab{}.
\newblock \showarticletitle{On the stability of fine-tuning bert:
  Misconceptions, explanations, and strong baselines}.
\newblock \bibinfo{journal}{\emph{arXiv preprint arXiv:2006.04884}}
  (\bibinfo{year}{2020}).
\newblock


\bibitem[\protect\citeauthoryear{Nguyen, Rosenberg, Song, Gao, Tiwary,
  Majumder, and Deng}{Nguyen et~al\mbox{.}}{2016}]%
        {nguyen2016ms_marco}
\bibfield{author}{\bibinfo{person}{Tri Nguyen}, \bibinfo{person}{Mir
  Rosenberg}, \bibinfo{person}{Xia Song}, \bibinfo{person}{Jianfeng Gao},
  \bibinfo{person}{Saurabh Tiwary}, \bibinfo{person}{Rangan Majumder}, {and}
  \bibinfo{person}{Li Deng}.} \bibinfo{year}{2016}\natexlab{}.
\newblock \showarticletitle{{MS MARCO}: A human generated machine reading
  comprehension dataset}.
\newblock \bibinfo{journal}{\emph{arXiv preprint arXiv:1611.09268}}
  (\bibinfo{year}{2016}).
\newblock


\bibitem[\protect\citeauthoryear{Nguyen, Stueker, Niehues, and Waibel}{Nguyen
  et~al\mbox{.}}{2020}]%
        {nguyen_2020_speech_aug}
\bibfield{author}{\bibinfo{person}{Thai-Son Nguyen}, \bibinfo{person}{Sebastian
  Stueker}, \bibinfo{person}{Jan Niehues}, {and} \bibinfo{person}{Alex
  Waibel}.} \bibinfo{year}{2020}\natexlab{}.
\newblock \showarticletitle{Improving sequence-to-sequence speech recognition
  training with on-the-fly data augmentation}. In
  \bibinfo{booktitle}{\emph{ICASSP 2020-2020 IEEE International Conference on
  Acoustics, Speech and Signal Processing (ICASSP)}}. IEEE,
  \bibinfo{pages}{7689--7693}.
\newblock


\bibitem[\protect\citeauthoryear{Nogueira and Cho}{Nogueira and Cho}{2019}]%
        {nogueira_prr_2019}
\bibfield{author}{\bibinfo{person}{Rodrigo Nogueira} {and}
  \bibinfo{person}{Kyunghyun Cho}.} \bibinfo{year}{2019}\natexlab{}.
\newblock \showarticletitle{Passage Re-ranking with {BERT}}.
\newblock \bibinfo{journal}{\emph{CoRR}}  \bibinfo{volume}{abs/1901.04085}
  (\bibinfo{year}{2019}).
\newblock
\urldef\tempurl%
\url{http://arxiv.org/abs/1901.04085}
\showURL{%
\tempurl}


\bibitem[\protect\citeauthoryear{Nugraha and Suyanto}{Nugraha and
  Suyanto}{2019}]%
        {nugraha_2019_typographic}
\bibfield{author}{\bibinfo{person}{Helmi~Satria Nugraha} {and}
  \bibinfo{person}{Suyanto Suyanto}.} \bibinfo{year}{2019}\natexlab{}.
\newblock \showarticletitle{Typographic-based data augmentation to improve a
  question retrieval in short dialogue system}. In
  \bibinfo{booktitle}{\emph{2019 International Seminar on Research of
  Information Technology and Intelligent Systems (ISRITI)}}. IEEE,
  \bibinfo{pages}{44--49}.
\newblock


\bibitem[\protect\citeauthoryear{Oord, Li, and Vinyals}{Oord
  et~al\mbox{.}}{2018}]%
        {oord2018representation}
\bibfield{author}{\bibinfo{person}{Aaron van~den Oord}, \bibinfo{person}{Yazhe
  Li}, {and} \bibinfo{person}{Oriol Vinyals}.} \bibinfo{year}{2018}\natexlab{}.
\newblock \showarticletitle{Representation learning with contrastive predictive
  coding}.
\newblock \bibinfo{journal}{\emph{arXiv preprint arXiv:1807.03748}}
  (\bibinfo{year}{2018}).
\newblock


\bibitem[\protect\citeauthoryear{Pasunuru, Celikyilmaz, Galley, Xiong, Zhang,
  Bansal, and Gao}{Pasunuru et~al\mbox{.}}{2021}]%
        {pasunuru2021data}
\bibfield{author}{\bibinfo{person}{Ramakanth Pasunuru}, \bibinfo{person}{Asli
  Celikyilmaz}, \bibinfo{person}{Michel Galley}, \bibinfo{person}{Chenyan
  Xiong}, \bibinfo{person}{Yizhe Zhang}, \bibinfo{person}{Mohit Bansal}, {and}
  \bibinfo{person}{Jianfeng Gao}.} \bibinfo{year}{2021}\natexlab{}.
\newblock \showarticletitle{Data augmentation for abstractive query-focused
  multi-document summarization}. In \bibinfo{booktitle}{\emph{Proceedings of
  the AAAI Conference on Artificial Intelligence}}, Vol.~\bibinfo{volume}{35}.
  \bibinfo{publisher}{Association for Computational Linguistics},
  \bibinfo{pages}{13666--13674}.
\newblock


\bibitem[\protect\citeauthoryear{Peng, Zhu, Zeng, and Gao}{Peng
  et~al\mbox{.}}{2020}]%
        {peng_2020_spoken_lang}
\bibfield{author}{\bibinfo{person}{Baolin Peng}, \bibinfo{person}{Chenguang
  Zhu}, \bibinfo{person}{Michael Zeng}, {and} \bibinfo{person}{Jianfeng Gao}.}
  \bibinfo{year}{2020}\natexlab{}.
\newblock \showarticletitle{Data augmentation for spoken language understanding
  via pretrained models}.
\newblock \bibinfo{journal}{\emph{arXiv e-prints}} (\bibinfo{year}{2020}),
  \bibinfo{pages}{arXiv--2004}.
\newblock


\bibitem[\protect\citeauthoryear{Raileanu, Goldstein, Yarats, Kostrikov, and
  Fergus}{Raileanu et~al\mbox{.}}{2020}]%
        {raileanu_2020_generalize_reinforce}
\bibfield{author}{\bibinfo{person}{Roberta Raileanu}, \bibinfo{person}{Max
  Goldstein}, \bibinfo{person}{Denis Yarats}, \bibinfo{person}{Ilya Kostrikov},
  {and} \bibinfo{person}{Rob Fergus}.} \bibinfo{year}{2020}\natexlab{}.
\newblock \showarticletitle{Automatic data augmentation for generalization in
  deep reinforcement learning}.
\newblock \bibinfo{journal}{\emph{arXiv preprint arXiv:2006.12862}}
  (\bibinfo{year}{2020}).
\newblock


\bibitem[\protect\citeauthoryear{Rogers, Kovaleva, and Rumshisky}{Rogers
  et~al\mbox{.}}{2020}]%
        {rogers2020primer}
\bibfield{author}{\bibinfo{person}{Anna Rogers}, \bibinfo{person}{Olga
  Kovaleva}, {and} \bibinfo{person}{Anna Rumshisky}.}
  \bibinfo{year}{2020}\natexlab{}.
\newblock \showarticletitle{A primer in bertology: What we know about how bert
  works}.
\newblock \bibinfo{journal}{\emph{Transactions of the Association for
  Computational Linguistics}}  \bibinfo{volume}{8} (\bibinfo{year}{2020}),
  \bibinfo{pages}{842--866}.
\newblock


\bibitem[\protect\citeauthoryear{Rudra and Anand}{Rudra and Anand}{2020}]%
        {rudra2020distant}
\bibfield{author}{\bibinfo{person}{Koustav Rudra} {and}
  \bibinfo{person}{Avishek Anand}.} \bibinfo{year}{2020}\natexlab{}.
\newblock \showarticletitle{Distant supervision in BERT-based adhoc document
  retrieval}. In \bibinfo{booktitle}{\emph{Proceedings of the 29th ACM
  International Conference on Information \& Knowledge Management}}.
  \bibinfo{pages}{2197--2200}.
\newblock


\bibitem[\protect\citeauthoryear{Sanh, Debut, Chaumond, and Wolf}{Sanh
  et~al\mbox{.}}{2019}]%
        {sanh2019distilbert}
\bibfield{author}{\bibinfo{person}{Victor Sanh}, \bibinfo{person}{Lysandre
  Debut}, \bibinfo{person}{Julien Chaumond}, {and} \bibinfo{person}{Thomas
  Wolf}.} \bibinfo{year}{2019}\natexlab{}.
\newblock \showarticletitle{DistilBERT, a distilled version of BERT: smaller,
  faster, cheaper and lighter}.
\newblock \bibinfo{journal}{\emph{arXiv preprint arXiv:1910.01108}}
  (\bibinfo{year}{2019}).
\newblock


\bibitem[\protect\citeauthoryear{Shorten and Khoshgoftaar}{Shorten and
  Khoshgoftaar}{2019}]%
        {shorten_2019_image_augmentation_survey}
\bibfield{author}{\bibinfo{person}{Connor Shorten} {and}
  \bibinfo{person}{Taghi~M Khoshgoftaar}.} \bibinfo{year}{2019}\natexlab{}.
\newblock \showarticletitle{A survey on image data augmentation for deep
  learning}.
\newblock \bibinfo{journal}{\emph{Journal of Big Data}} \bibinfo{volume}{6},
  \bibinfo{number}{1} (\bibinfo{year}{2019}), \bibinfo{pages}{1--48}.
\newblock


\bibitem[\protect\citeauthoryear{Shorten, Khoshgoftaar, and Furht}{Shorten
  et~al\mbox{.}}{2021}]%
        {shorten2021text}
\bibfield{author}{\bibinfo{person}{Connor Shorten}, \bibinfo{person}{Taghi~M
  Khoshgoftaar}, {and} \bibinfo{person}{Borko Furht}.}
  \bibinfo{year}{2021}\natexlab{}.
\newblock \showarticletitle{Text Data Augmentation for Deep Learning}.
\newblock \bibinfo{journal}{\emph{Journal of big Data}} \bibinfo{volume}{8},
  \bibinfo{number}{1} (\bibinfo{year}{2021}), \bibinfo{pages}{1--34}.
\newblock


\bibitem[\protect\citeauthoryear{Sohn}{Sohn}{2016}]%
        {sohn2016improved}
\bibfield{author}{\bibinfo{person}{Kihyuk Sohn}.}
  \bibinfo{year}{2016}\natexlab{}.
\newblock \showarticletitle{Improved deep metric learning with multi-class
  n-pair loss objective}. In \bibinfo{booktitle}{\emph{Advances in neural
  information processing systems}}. \bibinfo{pages}{1857--1865}.
\newblock


\bibitem[\protect\citeauthoryear{Strohman, Metzler, Turtle, and Croft}{Strohman
  et~al\mbox{.}}{2005}]%
        {strohman_indri_2005}
\bibfield{author}{\bibinfo{person}{Trevor Strohman}, \bibinfo{person}{Donald
  Metzler}, \bibinfo{person}{Howard Turtle}, {and} \bibinfo{person}{W~Bruce
  Croft}.} \bibinfo{year}{2005}\natexlab{}.
\newblock \showarticletitle{Indri: A language model-based search engine for
  complex queries}. In \bibinfo{booktitle}{\emph{Proceedings of the
  International Conference on Intelligent Analysis}}, Vol.~\bibinfo{volume}{2}.
  \bibinfo{pages}{2--6}.
\newblock


\bibitem[\protect\citeauthoryear{Sun, Xia, Yin, Liang, Yu, and He}{Sun
  et~al\mbox{.}}{2020}]%
        {sun_2020_mixup}
\bibfield{author}{\bibinfo{person}{Lichao Sun}, \bibinfo{person}{Congying Xia},
  \bibinfo{person}{Wenpeng Yin}, \bibinfo{person}{Tingting Liang},
  \bibinfo{person}{Philip~S Yu}, {and} \bibinfo{person}{Lifang He}.}
  \bibinfo{year}{2020}\natexlab{}.
\newblock \showarticletitle{Mixup-Transformer: Dynamic Data Augmentation for
  NLP Tasks}.
\newblock \bibinfo{journal}{\emph{arXiv preprint arXiv:2010.02394}}
  (\bibinfo{year}{2020}).
\newblock


\bibitem[\protect\citeauthoryear{Van, Yadav, and Surdeanu}{Van
  et~al\mbox{.}}{2021}]%
        {van_2021_machine_reading}
\bibfield{author}{\bibinfo{person}{Hoang Van}, \bibinfo{person}{Vikas Yadav},
  {and} \bibinfo{person}{Mihai Surdeanu}.} \bibinfo{year}{2021}\natexlab{}.
\newblock \showarticletitle{Cheap and Good? Simple and Effective Data
  Augmentation for Low Resource Machine Reading}.
\newblock \bibinfo{journal}{\emph{arXiv preprint arXiv:2106.04134}}
  (\bibinfo{year}{2021}).
\newblock


\bibitem[\protect\citeauthoryear{Voorhees}{Voorhees}{2005}]%
        {robust04}
\bibfield{author}{\bibinfo{person}{Ellen~M. Voorhees}.}
  \bibinfo{year}{2005}\natexlab{}.
\newblock \showarticletitle{Overview of the {TREC} 2004 robust track}. In
  \bibinfo{booktitle}{\emph{TREC, volume Special Publication 500-261}}.
  \bibinfo{publisher}{National Institute of Standards and Technology (NIST)}.
\newblock


\bibitem[\protect\citeauthoryear{Wadden, Lin, Lo, Wang, van Zuylen, Cohan, and
  Hajishirzi}{Wadden et~al\mbox{.}}{2020}]%
        {Wadden2020FactOF}
\bibfield{author}{\bibinfo{person}{David Wadden}, \bibinfo{person}{Shanchuan
  Lin}, \bibinfo{person}{Kyle Lo}, \bibinfo{person}{Lucy~Lu Wang},
  \bibinfo{person}{Madeleine van Zuylen}, \bibinfo{person}{Arman Cohan}, {and}
  \bibinfo{person}{Hannaneh Hajishirzi}.} \bibinfo{year}{2020}\natexlab{}.
\newblock \showarticletitle{Fact or Fiction: Verifying Scientific Claims}. In
  \bibinfo{booktitle}{\emph{EMNLP}}.
\newblock


\bibitem[\protect\citeauthoryear{Wang and Isola}{Wang and Isola}{2020}]%
        {wang2020understanding}
\bibfield{author}{\bibinfo{person}{Tongzhou Wang} {and}
  \bibinfo{person}{Phillip Isola}.} \bibinfo{year}{2020}\natexlab{}.
\newblock \showarticletitle{Understanding contrastive representation learning
  through alignment and uniformity on the hypersphere}. In
  \bibinfo{booktitle}{\emph{International Conference on Machine Learning}}.
  PMLR, \bibinfo{pages}{9929--9939}.
\newblock


\bibitem[\protect\citeauthoryear{Wu, Xiong, Yu, and Lin}{Wu
  et~al\mbox{.}}{2018}]%
        {wu2018unsupervised}
\bibfield{author}{\bibinfo{person}{Zhirong Wu}, \bibinfo{person}{Yuanjun
  Xiong}, \bibinfo{person}{Stella~X Yu}, {and} \bibinfo{person}{Dahua Lin}.}
  \bibinfo{year}{2018}\natexlab{}.
\newblock \showarticletitle{Unsupervised feature learning via non-parametric
  instance discrimination}. In \bibinfo{booktitle}{\emph{Proceedings of the
  IEEE conference on computer vision and pattern recognition}}.
  \bibinfo{pages}{3733--3742}.
\newblock


\bibitem[\protect\citeauthoryear{Xiong, Xiong, Li, Tang, Liu, Bennett, Ahmed,
  and Overwijk}{Xiong et~al\mbox{.}}{2020}]%
        {xiong2020approximate}
\bibfield{author}{\bibinfo{person}{Lee Xiong}, \bibinfo{person}{Chenyan Xiong},
  \bibinfo{person}{Ye Li}, \bibinfo{person}{Kwok-Fung Tang},
  \bibinfo{person}{Jialin Liu}, \bibinfo{person}{Paul Bennett},
  \bibinfo{person}{Junaid Ahmed}, {and} \bibinfo{person}{Arnold Overwijk}.}
  \bibinfo{year}{2020}\natexlab{}.
\newblock \showarticletitle{Approximate Nearest Neighbor Negative Contrastive
  Learning for Dense Text Retrieval}.
\newblock \bibinfo{journal}{\emph{arXiv preprint arXiv:2007.00808}}
  (\bibinfo{year}{2020}).
\newblock


\bibitem[\protect\citeauthoryear{Xu, Li, and Zhu}{Xu et~al\mbox{.}}{2020}]%
        {xu_2020_image_segment}
\bibfield{author}{\bibinfo{person}{Ju Xu}, \bibinfo{person}{Mengzhang Li},
  {and} \bibinfo{person}{Zhanxing Zhu}.} \bibinfo{year}{2020}\natexlab{}.
\newblock \showarticletitle{Automatic data augmentation for 3D medical image
  segmentation}. In \bibinfo{booktitle}{\emph{International Conference on
  Medical Image Computing and Computer-Assisted Intervention}}. Springer,
  \bibinfo{pages}{378--387}.
\newblock


\bibitem[\protect\citeauthoryear{Yanagi, Togo, Ogawa, and Haseyama}{Yanagi
  et~al\mbox{.}}{2020}]%
        {yanagi_2020_image}
\bibfield{author}{\bibinfo{person}{Rintaro Yanagi}, \bibinfo{person}{Ren Togo},
  \bibinfo{person}{Takahiro Ogawa}, {and} \bibinfo{person}{Miki Haseyama}.}
  \bibinfo{year}{2020}\natexlab{}.
\newblock \showarticletitle{Image Retrieval with Data Augmentation of Sentence
  Labels Based on Paraphrasing}. In \bibinfo{booktitle}{\emph{2020 IEEE
  International Conference on Consumer Electronics-Taiwan (ICCE-Taiwan)}}.
  IEEE, \bibinfo{pages}{1--2}.
\newblock


\bibitem[\protect\citeauthoryear{Yang, Wei, Jiao, Jiang, and Yang}{Yang
  et~al\mbox{.}}{2021}]%
        {yang2021xmoco}
\bibfield{author}{\bibinfo{person}{Nan Yang}, \bibinfo{person}{Furu Wei},
  \bibinfo{person}{Binxing Jiao}, \bibinfo{person}{Daxing Jiang}, {and}
  \bibinfo{person}{Linjun Yang}.} \bibinfo{year}{2021}\natexlab{}.
\newblock \showarticletitle{xmoco: Cross momentum contrastive learning for
  open-domain question answering}. In \bibinfo{booktitle}{\emph{Proceedings of
  the 59th Annual Meeting of the Association for Computational Linguistics and
  the 11th International Joint Conference on Natural Language Processing
  (Volume 1: Long Papers)}}. \bibinfo{pages}{6120--6129}.
\newblock


\bibitem[\protect\citeauthoryear{Yang, Xie, Tan, Xiong, Li, and Lin}{Yang
  et~al\mbox{.}}{2019}]%
        {yang_2019_data}
\bibfield{author}{\bibinfo{person}{Wei Yang}, \bibinfo{person}{Yuqing Xie},
  \bibinfo{person}{Luchen Tan}, \bibinfo{person}{Kun Xiong},
  \bibinfo{person}{Ming Li}, {and} \bibinfo{person}{Jimmy Lin}.}
  \bibinfo{year}{2019}\natexlab{}.
\newblock \showarticletitle{Data augmentation for bert fine-tuning in
  open-domain question answering}.
\newblock \bibinfo{journal}{\emph{arXiv preprint arXiv:1904.06652}}
  (\bibinfo{year}{2019}).
\newblock


\bibitem[\protect\citeauthoryear{Yang, Jin, Lin, Guo, and Cer}{Yang
  et~al\mbox{.}}{2020}]%
        {yang_2020_cross_attention}
\bibfield{author}{\bibinfo{person}{Yinfei Yang}, \bibinfo{person}{Ning Jin},
  \bibinfo{person}{Kuo Lin}, \bibinfo{person}{Mandy Guo}, {and}
  \bibinfo{person}{Daniel Cer}.} \bibinfo{year}{2020}\natexlab{}.
\newblock \showarticletitle{Neural Retrieval for Question Answering with
  Cross-Attention Supervised Data Augmentation}.
\newblock \bibinfo{journal}{\emph{arXiv preprint arXiv:2009.13815}}
  (\bibinfo{year}{2020}).
\newblock


\bibitem[\protect\citeauthoryear{Yao, Yang, Zhang, Chen, and Luo}{Yao
  et~al\mbox{.}}{2020}]%
        {yao_2020_domain}
\bibfield{author}{\bibinfo{person}{Liang Yao}, \bibinfo{person}{Baosong Yang},
  \bibinfo{person}{Haibo Zhang}, \bibinfo{person}{Boxing Chen}, {and}
  \bibinfo{person}{Weihua Luo}.} \bibinfo{year}{2020}\natexlab{}.
\newblock \showarticletitle{Domain transfer based data augmentation for neural
  query translation}. In \bibinfo{booktitle}{\emph{Proceedings of the 28th
  International Conference on Computational Linguistics}}.
  \bibinfo{pages}{4521--4533}.
\newblock


\bibitem[\protect\citeauthoryear{Yilmaz, Yang, Zhang, and Lin}{Yilmaz
  et~al\mbox{.}}{2019}]%
        {yilmaz2019cross}
\bibfield{author}{\bibinfo{person}{Zeynep~Akkalyoncu Yilmaz},
  \bibinfo{person}{Wei Yang}, \bibinfo{person}{Haotian Zhang}, {and}
  \bibinfo{person}{Jimmy Lin}.} \bibinfo{year}{2019}\natexlab{}.
\newblock \showarticletitle{Cross-domain modeling of sentence-level evidence
  for document retrieval}. In \bibinfo{booktitle}{\emph{Proceedings of the 2019
  conference on empirical methods in natural language processing and the 9th
  international joint conference on natural language processing
  (EMNLP-IJCNLP)}}. \bibinfo{pages}{3490--3496}.
\newblock


\bibitem[\protect\citeauthoryear{Yu, Yang, Liu, Li, Zhang, and Zhao}{Yu
  et~al\mbox{.}}{2019}]%
        {yu_2019_hierarchical}
\bibfield{author}{\bibinfo{person}{Shujuan Yu}, \bibinfo{person}{Jie Yang},
  \bibinfo{person}{Danlei Liu}, \bibinfo{person}{Runqi Li},
  \bibinfo{person}{Yun Zhang}, {and} \bibinfo{person}{Shengmei Zhao}.}
  \bibinfo{year}{2019}\natexlab{}.
\newblock \showarticletitle{Hierarchical data augmentation and the application
  in text classification}.
\newblock \bibinfo{journal}{\emph{IEEE Access}}  \bibinfo{volume}{7}
  (\bibinfo{year}{2019}), \bibinfo{pages}{185476--185485}.
\newblock


\bibitem[\protect\citeauthoryear{Zeng, Qiu, Memmi, and Qiu}{Zeng
  et~al\mbox{.}}{2020}]%
        {zeng_2020_adversarial}
\bibfield{author}{\bibinfo{person}{Yi Zeng}, \bibinfo{person}{Han Qiu},
  \bibinfo{person}{Gerard Memmi}, {and} \bibinfo{person}{Meikang Qiu}.}
  \bibinfo{year}{2020}\natexlab{}.
\newblock \showarticletitle{A data augmentation-based defense method against
  adversarial attacks in neural networks}. In
  \bibinfo{booktitle}{\emph{International Conference on Algorithms and
  Architectures for Parallel Processing}}. Springer, \bibinfo{pages}{274--289}.
\newblock


\bibitem[\protect\citeauthoryear{Zhang, Zhao, and LeCun}{Zhang
  et~al\mbox{.}}{2015}]%
        {zhang2015character}
\bibfield{author}{\bibinfo{person}{Xiang Zhang}, \bibinfo{person}{Junbo Zhao},
  {and} \bibinfo{person}{Yann LeCun}.} \bibinfo{year}{2015}\natexlab{}.
\newblock \showarticletitle{Character-level convolutional networks for text
  classification}.
\newblock \bibinfo{journal}{\emph{Advances in neural information processing
  systems}}  \bibinfo{volume}{28} (\bibinfo{year}{2015}),
  \bibinfo{pages}{649--657}.
\newblock


\bibitem[\protect\citeauthoryear{Zhang and Sabuncu}{Zhang and Sabuncu}{2018}]%
        {zhang2018generalized::augprob}
\bibfield{author}{\bibinfo{person}{Zhilu Zhang} {and} \bibinfo{person}{Mert~R
  Sabuncu}.} \bibinfo{year}{2018}\natexlab{}.
\newblock \showarticletitle{Generalized cross entropy loss for training deep
  neural networks with noisy labels}. In \bibinfo{booktitle}{\emph{32nd
  Conference on Neural Information Processing Systems (NeurIPS)}}.
\newblock


\bibitem[\protect\citeauthoryear{Zhong, Zheng, Kang, Li, and Yang}{Zhong
  et~al\mbox{.}}{2020}]%
        {zhong_2020_erasing}
\bibfield{author}{\bibinfo{person}{Zhun Zhong}, \bibinfo{person}{Liang Zheng},
  \bibinfo{person}{Guoliang Kang}, \bibinfo{person}{Shaozi Li}, {and}
  \bibinfo{person}{Yi Yang}.} \bibinfo{year}{2020}\natexlab{}.
\newblock \showarticletitle{Random erasing data augmentation}. In
  \bibinfo{booktitle}{\emph{Proceedings of the AAAI Conference on Artificial
  Intelligence}}, Vol.~\bibinfo{volume}{34}. \bibinfo{pages}{13001--13008}.
\newblock


\bibitem[\protect\citeauthoryear{Zhu, Wang, Chen, and Liu}{Zhu
  et~al\mbox{.}}{2020}]%
        {zhu_2020_dialog}
\bibfield{author}{\bibinfo{person}{Qingqing Zhu}, \bibinfo{person}{Xiwei Wang},
  \bibinfo{person}{Chen Chen}, {and} \bibinfo{person}{Junfei Liu}.}
  \bibinfo{year}{2020}\natexlab{}.
\newblock \showarticletitle{Data Augmentation for Retrieval-and
  Generation-Based Dialog Systems}. In \bibinfo{booktitle}{\emph{2020 IEEE 6th
  International Conference on Computer and Communications (ICCC)}}. IEEE,
  \bibinfo{pages}{1716--1720}.
\newblock


\bibitem[\protect\citeauthoryear{Zoph, Cubuk, Ghiasi, Lin, Shlens, and Le}{Zoph
  et~al\mbox{.}}{2020}]%
        {zoph_2020_object_detection}
\bibfield{author}{\bibinfo{person}{Barret Zoph}, \bibinfo{person}{Ekin~D
  Cubuk}, \bibinfo{person}{Golnaz Ghiasi}, \bibinfo{person}{Tsung-Yi Lin},
  \bibinfo{person}{Jonathon Shlens}, {and} \bibinfo{person}{Quoc~V Le}.}
  \bibinfo{year}{2020}\natexlab{}.
\newblock \showarticletitle{Learning data augmentation strategies for object
  detection}. In \bibinfo{booktitle}{\emph{European Conference on Computer
  Vision}}. Springer, \bibinfo{pages}{566--583}.
\newblock


\end{thebibliography}
